\documentclass[11pt]{article}

\usepackage[margin=1in]{geometry}
\usepackage[T1]{fontenc}
\usepackage[utf8]{inputenc}
\usepackage{newtxtext,newtxmath}
\usepackage{amsmath}
\usepackage{graphicx}
\usepackage{booktabs}
\usepackage{caption}
\usepackage[numbers,sort&compress]{natbib}
\usepackage{bibunits}
\usepackage{setspace}
\usepackage[hidelinks]{hyperref}
\usepackage{siunitx}
\usepackage{authblk}
\usepackage[section]{placeins}

\hypersetup{
  pdftitle={Mind-wandering preserves word-level coupling while disrupting comprehension},
  pdfauthor={Anirudh Natarajan and Paul Sajda},
  pdfsubject={Mind-wandering, lexical guidance, comprehension and EEG},
  pdfkeywords={mind-wandering; reading; eye movements; alpha; EEG}
}
\defaultbibliographystyle{unsrtnat}
\defaultbibliography{refs}
\makeatletter
\newcommand{\maincitations}{\def\@extra@b@citeb{main}\def\@extra@binfo{main}}
\newcommand{\suppcitations}{\def\@extra@b@citeb{supp}\def\@extra@binfo{supp}}
\makeatother
\graphicspath{{figs/}}
\title{\bfseries Mind-wandering preserves word-level coupling\\
while disrupting comprehension}

\author[1]{Anirudh Natarajan\thanks{Correspondence: \texttt{amn2225@columbia.edu}}}
\author[1,2,3,4]{Paul Sajda}
\affil[1]{Department of Biomedical Engineering, Columbia University, New York, NY 10027, USA}
\affil[2]{Department of Electrical Engineering, Columbia University, New York, NY 10027, USA}
\affil[3]{Department of Radiology, Columbia University Irving Medical Center, New York, NY 10032, USA}
\affil[4]{The Data Science Institute, Columbia University, New York, NY 10027, USA}
\date{}

\begin{document}
\maketitle
\maincitations
\begin{bibunit}[unsrtnat]

\begin{abstract}
\noindent
Readers can move their eyes through a page but retain little of its meaning when their minds wander. We tested whether word properties still guide their eyes during these lapses. We analyzed public eye-tracking and electroencephalography (EEG) data from 44 adults who read articles. They marked the text read during each lapse they noticed. We linked these spans to the sentences needed for later comprehension questions. Word frequency, length and surprisal still predicted which words readers skipped and how long they looked at them. Compared with on-task reading, these effects met equivalence bounds of $\pm$10\% for skipping and $\pm$20\% for gaze duration. Readers spent 14.6\% longer on words during lapses. Yet readers answered less accurately when lapses covered sentences needed for a later question. The adjusted difference was 25.7 percentage points for lapses covering all versus none of these sentences. The fixation-related EEG response was 42.7\% smaller during lapses. At left posterior-lateral sensor P7, the median increase in alpha power was 10.0\%. Occipital sensor Oz showed no reliable increase. These results suggest that word properties can guide the eyes while readers fail to form or retain a useful representation of the text.
\end{abstract}

\vspace{0.5em}

\newpage

\section{Introduction}

Readers sometimes reach the bottom of a page having retained little of what they read. Their eyes moved through the text, but successful viewing did not ensure comprehension. Mind-wandering is associated with poorer comprehension and altered eye movements \citep{smallwood2006,schooler2011,bonifacci2022,meziere2025}. What remains uncertain is whether the words continue to guide the eyes during these lapses, and how that guidance relates to the information subsequently lost.

\begin{figure}[!tbp]
\centering
\includegraphics[width=\textwidth]{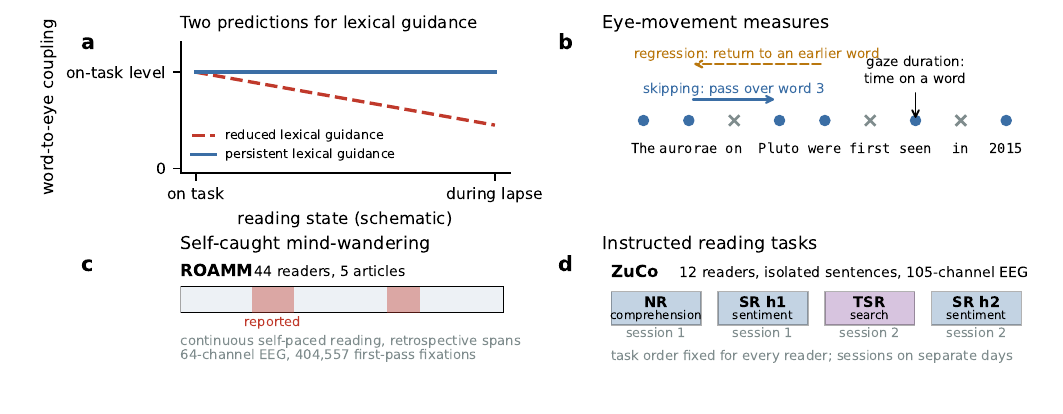}
\caption{\textbf{Question and design.}
(a) Two accounts make different predictions for lexical guidance during a lapse: reduced influence of word properties on eye movements, or persistent influence despite impaired comprehension. Lines are schematic predictions, not fitted lapse-depth functions.
(b) Skipping, gaze duration and returns to earlier words measure different aspects of eye-movement control. Circles mark fixated words and crosses mark words the reader stepped over.
(c) The primary sample, 44 readers. Readers read connected articles at their own pace, reported
each lapse as they caught it, and marked the span of text it had covered.
(d) The independent task sample, 12 readers. Readers read isolated sentences for comprehension, sentiment judgments or semantic-relation search. Task order was the same for every reader, and the two encyclopedic tasks were in different
sessions. Section~\ref{sec:goal} uses this structure.}
\label{fig:design}
\end{figure}

We compare two explanations for this failure (Fig.~\ref{fig:design}a). Under an explanation requiring reduced lexical guidance, mind-wandering weakens the influence of word processing on eye-movement control. This prediction follows from a reading-specific formulation of perceptual decoupling: the products of word identification become less available to the systems that program eye movements \citep{smallwood2011cascade,schooler2011}. Earlier studies reported reduced lexical influences on fixation durations before self-caught mindless reading, with attenuation varying with lapse depth \citep{reichle2010,schad2012}. An alternative account allows lexical guidance to continue while the construction or retention of a usable text representation becomes less effective. Research connecting mind-wandering to impaired situation-model construction motivates this possibility \citep{smallwood2008attention}. The explanations differ in whether they predict weaker word-to-eye coupling.

The distinction requires more than a comparison of average fixation durations or skipping rates. Eye-movement models distinguish where to look from when to move on, with lexical information influencing skipping, fixation timing and subsequent returns \citep{reichle2003,engbert2005,rayner2009}. A reader can spend longer on all words while keeping the difference between rare and common words. Conversely, more skipping need not imply weaker lexical guidance. We therefore measured the influence of word frequency, length and contextual surprisal on skipping and gaze duration, supplemented by refixation, regression and model-based analyses. We tested whether changes fell within specified equivalence bounds: 10\% for skipping and 20\% for gaze duration.

Neural state changes need not imply a loss of this guidance either. Earlier work reports attenuated early potentials during reading, altered surprisal responses, and reduced brain--language alignment \citep{broadway2015,southwell2026eeg,sun2026alignment}. Alpha power also increases during mind-wandering in other tasks \citep{compton2019}. We distinguish changes in the response common to fixations from changes in its modulation by word properties. We also distinguish a left posterior-lateral alpha effect from a general occipital increase. Scalp EEG can test whether neural changes accompany persistent coupling, but cannot by itself identify their cortical source or explain comprehension failure.

The ROAMM dataset links self-marked lapse spans, eye movements, EEG and a comprehension question for each page \citep{roammdata}. This permits three tests: whether lexical guidance persists during reported lapses; whether those lapses overlap the content required by a later question; and whether neural state measures predict accuracy when measured on that content. An independent sample performing semantic-relation search provides a sensitivity check for the duration measure \citep{hollenstein2018}. Together, the analyses test whether sustained, self-caught mind-wandering requires reduced lexical guidance or permits a separation between word-level eye-movement control and successful comprehension.

\section{Methods}

\subsection{Participants, task and lapse reports}

We analyzed two public datasets; full procedures are in SI~S16. In ROAMM, 44 adults read five Wikipedia articles, each divided into ten pages, with simultaneous binocular eye tracking (1000\,Hz) and 64-channel EEG (2048\,Hz; released at 256\,Hz) \citep{roammdata}. Reading was self-paced. Readers pressed a key when they caught themselves mind-wandering and marked its onset and offset in the text \citep{remind2026}. We inherited the providers' labels: onset is the first fixation on the marked onset word, and offset is 2\,s before the report. The final 2\,s before reports and page advances were excluded. Only first-pass reading before the first report or page advance was labeled, yielding 404,557 fixations; 7.2\% were inside a reported lapse. After each article, readers answered one four-option question per page. These self-caught reports sample lapses that reached awareness.

The independent ZuCo sample comprised 12 adults reading isolated sentences for comprehension, semantic-relation search or sentiment judgments \citep{hollenstein2018}. Because task order was fixed, we estimated the search contrast after subtracting the change between the sentiment-task halves in the two sessions (SI~S10). This comparison checks sensitivity to a different reading goal; it is not a randomized manipulation of mind-wandering.

\subsection{Word properties, eye movements and inference}

Word predictors were frequency (\texttt{wordfreq} Zipf value), length in letters and GPT-2 contextual surprisal in bits \citep{speer2022wordfreq,radford2019}. Surprisal used the preceding words of the sentence; sentence-initial words without a value were excluded from surprisal contrasts. Skipping required a demonstrated forward step over a word, primarily within a gap of at most four intervening words. We also used a purely layout-based rule and the standard first-pass definition, which counts a word as skipped even if it is fixated later (SI~S4). Gaze duration summed first-pass fixations on a word. We separately measured immediate refixations, regressions and returns to skipped words; these events do not establish successful repair.

We use \emph{coupling} for the measured association between a word property and an eye-movement measure; \emph{lexical guidance} is its interpretation in reading. Skipping and return coupling used Somers' $D$; gaze-duration coupling used slopes on log duration. A model with reader and corpus-token fixed effects compared the same corpus words read in different states. Reader was the unit of inference. We used reader-level bootstrap intervals for paired contrasts and normal-Wald intervals from reader-clustered standard errors for the fixed-effects model. We corrected families of three word-property tests using Holm's procedure and tested equivalence within 10\% and 20\% of the on-task association \citep{lakens2017}. SI~S15 reports the bounds, detectable changes and sample sizes for each contrast. A network-based test evaluated linguistic information beyond the selected associations (SI~S5).

\subsection{EEG responses and ongoing alpha}
\label{sec:methods-eeg}
\label{sec:methods-alpha}

We used the providers' cleaned EEG and estimated overlapping fixation-related responses with linear deconvolution \citep{dimigen2021,ehinger2019}. The design contained an intercept, word properties, fixation duration, lapse state and state interactions with frequency and surprisal. Ridge penalties were proportional to each design column's energy; sensitivity to the penalty is reported in SI~S7. We assessed frequency over occipitotemporal sensors at 150--290\,ms and surprisal over centroparietal sensors at 300--450\,ms. A leave-one-reader-out factor tested proportional rescaling of the intercept field. Incoming-saccade covariates and amplitude-matched samples checked ocular explanations.

Ongoing power was estimated in nonoverlapping 2\,s segments wholly inside one state. We tested channels with a reader-level sign-flip max-$t$ procedure controlling family-wise error over 64 sensors. The clean contrast excluded blink segments and the 10\,s following lapse offset, covering report and annotation periods. We repeated the analysis after subtracting fixation-evoked activity (SI~S9). The main scalp map shows this clean contrast; per-reader P7 and Oz values show the unexcluded contrast and are identified as such.

\subsection{Comprehension and content specificity}
\label{sec:methods-comp}

Answer-bearing sentences were annotated before outcome modeling, with an independent automatic annotation and answerability checks (SI~S12). Trials required at least three fixations on the answer span and ten elsewhere, retaining 1804 of 2200 outcomes. Linear probability models absorbed reader and item effects, with reader-clustered standard errors and normal-Wald confidence intervals. We compared the answer-span association against 1000 random text sets from the same page, each matching the answer span in word count, adjusting for page coverage. We also tested gaze coverage and fixation-aligned alpha on the answer span. Alpha was mean log 8--13\,Hz power from 500\,ms before to 500\,ms after fixation onset, standardized within run and averaged over answer-span fixations.

The study was not registered. Duration was analyzed first; hypotheses and rules for the subsequent skipping and return analyses were documented before their contrasts were computed. The skipping audit, neural rescaling, alpha and alignment analyses followed earlier findings. The duration equivalence tests were calculated from the saved estimates during manuscript preparation; they were not preregistered. SI~S16 documents the analysis sequence, exclusions and software.

\section{Results}

\subsection{On-task associations and reported lapse spans}
\label{sec:instrument}

During on-task reading, all three eye-movement decisions and both deconvolved kernels changed with
word properties in the expected way: rarer, longer and more surprising words were fixated longer,
skipped less and returned to more, and the kernels showed an occipitotemporal frequency response
and a centroparietal N400 (all $p<10^{-5}$; SI~S1, Fig.~\ref{fig:instrument}).

The reported spans also showed the expected properties of mind-wandering. All 44 readers marked at
least one span, and per-reader rates ranged from 0.2\% to 24.2\% of fixations. Fixations inside a
reported span were 5.0\% longer than the same reader's on-task fixations (95\% CI $[3.3, 6.6]$,
$p=7.9\times10^{-7}$, 39 of 44 readers). The results below do not depend on the exact placement of the spans: shifting each reader's labels
by up to 15 words, or cutting spans into position bins, left the skipping-coupling estimates unchanged
(SI~S2).

\subsection{Lexical influences on skipping and gaze duration persist}
\label{sec:preserved}

\begin{figure}[!tbp]
\centering
\includegraphics[width=\textwidth]{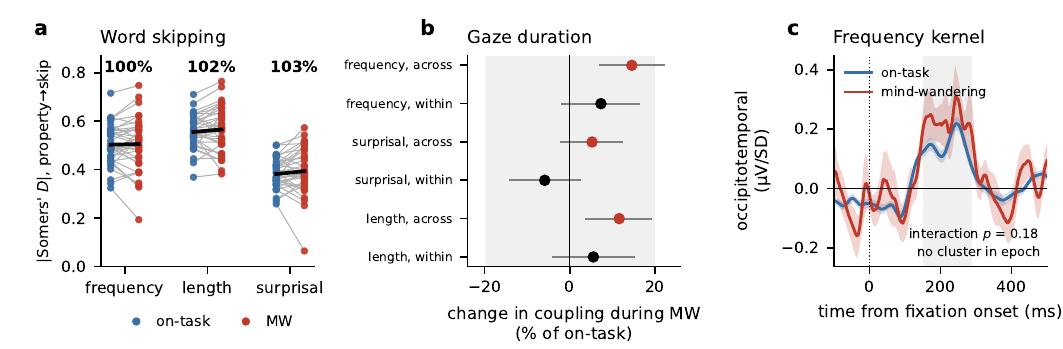}
\caption{\textbf{Lexical influences on eye movements persist; the neural frequency contrast is less precise.}
(a) Absolute Somers' $D$ ($|D|$) for word properties and skipping, per reader and state. The text reports signed $D$: frequency has a positive association, whereas length and surprisal have negative associations. A more negative $\Delta D$ for length or surprisal therefore means stronger coupling. Grey lines join the two states within a reader, black
lines show the condition means, and percentages give retention. The two lowest mind-wandering points are the same reader, the one
with the fewest mind-wandering words (188); excluding them moves retention to 101.2\% and
104.6\%.
(b) Change in the gaze-duration slope, across words (colored) and within token instance with
reader and token fixed effects (black); only the within-token estimate is identified. Bars are 95\% intervals; the shaded band is the
$\pm$20\% equivalence region. Within-token intervals are normal-Wald intervals from reader-clustered standard errors; equivalence tests use $t_{43}$.
(c) Deconvolved occipitotemporal frequency kernel by state. The mind-wandering kernel is noisier because it uses only one fourteenth of the fixations. Its point
estimate is larger, and the interaction is not reliable. Readers per panel differ with the words each contrast requires (36 in a, 44 in b and c).
Return contrasts, the surprisal kernel and the precision of each estimate are in Table~\ref{tab:equiv},
SI~S3 and SI~S6.}
\label{fig:preserved}
\end{figure}

We first tested whether word properties had weaker associations with skipping and gaze duration inside reported lapses (Fig.~\ref{fig:preserved}). Equivalence tests supported changes within 10\% for skipping and 20\% for gaze duration. These bounds describe different levels of precision.

\paragraph{Word skipping.} A word counted as skipped only if the reader's scan path stepped over it
in a single forward saccade spanning at most four intervening words. Words inside larger
positional gaps were excluded rather than scored, as those gaps are almost entirely return
sweeps (Section~\ref{sec:skip}). This left 332{,}016 words from 36 readers with at least 150 words in each state. Of these words,
21{,}633 fell inside a reported span. Retention of the on-task
association was 100.5\% for frequency (mind-wandering minus on-task $\Delta D=+0.0025$, 95\% CI
$[-0.0167, +0.0214]$, $p=0.80$), 101.8\% for length ($\Delta D=-0.0100$, $[-0.0283, +0.0099]$,
$p=0.32$) and 103.1\% for surprisal ($\Delta D=-0.0118$, $[-0.0326, +0.0099]$, $p=0.30$). All
three were statistically equivalent within $\pm$10\% of the on-task effect (TOST
$p=1.1\times10^{-5}$, $2.6\times10^{-5}$ and $0.011$; BF$_{01}=5.4$, $3.5$ and $3.3$). The
standard first-pass definition gives the same result when steps stay within a text line
(retention 98.4\%, 101.8\% and 99.6\%; TOST $p\le0.006$; SI~S4).

The 10\% bound states the precision of the measure, not a prediction from earlier work: the
attenuations reported before self-caught lapses were measured on fixation durations
\citep{reichle2010,schad2012}, and we know of no earlier estimate for skipping.
Section~\ref{sec:goal} shows what this null can and cannot establish on its own.

The skipping result held across gap thresholds, fully off-task reports, run-position bins, article exclusions and span-preserving controls (SI~S3).

\paragraph{Refixations and regressions.} Refixation coupling held at 96\% to 99\% of its on-task value (35 readers, all
$p>0.5$). Regression targeting showed no reliable change but is too imprecise to bound anything,
and corrective-return coupling could be estimated in only 16 readers (Table~\ref{tab:equiv},
SI~S3).

\paragraph{Gaze duration.} Changes in gaze-duration coupling fell within the 20\% bound. We compared the same words read in the two states. Different readers read 7{,}264 of 10{,}183
corpus tokens (71.3\%) both on task and during mind-wandering. We therefore absorbed token and
reader identity in the fixed-effects model. The estimated interactions were: $+7.3\%$
of the on-task slope for frequency (95\% CI $[-1.9, +16.6]$, $p=0.13$), $-5.9\%$ for surprisal
($[-14.4, +2.6]$, $p=0.18$) and $+5.5\%$ for length ($[-4.2, +15.2]$, $p=0.27$; $n=234{,}800$).
All three interactions passed TOST at $\pm$20\% of the observed on-task slope: frequency $p=0.0051$, surprisal $p=0.0011$ and length $p=0.0027$ (Holm-adjusted $p\le0.0055$). None passed at $\pm$10\% ($p\ge0.17$). Duration coupling is therefore bounded less tightly than skipping coupling. The tests use the saved reader-clustered standard errors and $t_{43}$ reference distribution, treating the observed on-task slopes as fixed reference values (SI~S15).

The adjusted state coefficient corresponds to 10.8\% longer gaze duration
($\beta=+0.103$, SE $=0.011$, $p=1.6\times10^{-11}$). Compared across words, the mind-wandering slopes were slightly larger than on task (114.6\% for
frequency, $p=0.001$; 111.6\% for length, $p=0.007$). That comparison can reflect differences in the words encountered during each state, because those words were not randomly assigned.

\paragraph{Model-based coupling.} Each test above fixes a decision and a word property in
advance. To remove that choice, a network trained on on-task fixations predicted the next fixated
word and the current fixation duration from page geometry, reading history and word properties
\citep{deng2023eyettention,bolliger2025scandl,hahn2023task}. During mind-wandering the
word-property information retained $0.911$ of its on-task value for target choice (95\% CI
$[0.822, 0.995]$) and $1.071$ for duration. We do not present the 9\% shortfall as established,
because it is not significant under a span-preserving permutation ($p=0.067$; SI~S5).

\paragraph{The deconvolved kernels.} The frequency and surprisal estimates differ in precision. The frequency-kernel estimate did not show a reliable attenuation. Its interaction with mind-wandering was
$+0.083$\,\si{\micro\volt} per SD ($p=0.18$, $n=44$), a change of $+56.5\%$ of the on-task effect with a
95\% interval of $[-14.3, +143.5]\%$. The interval is wide, and the smallest change detectable at
80\% power is 118\% of the on-task effect. The decoupling account predicts a reduction, and the one-sided bound excludes attenuations greater
than 4\%. It does so only because the estimate is on the side of an increase, not
because the contrast is precise (BF$_{01}=2.5$).

The surprisal kernel could not be tested usefully. Its change was $-35.4\%$ of the on-task effect ($p=0.72$) with a 95\% interval of $[-234.6,
+151.5]\%$, and the smallest detectable change is 283\%. The
test cannot distinguish a preserved N400 from an abolished one, and neither a whole-epoch cluster
test nor a pooled encoding model resolves it (SI~S6).

The tightest constraints concern skipping and gaze duration. The return measures are less informative, and the neural contrasts do not establish intact semantic processing (SI~S15, Table~\ref{tab:equiv}).

\subsection{Lapse overlap with answer-bearing text predicts accuracy}
\label{sec:comp}

\begin{figure}[!tbp]
\centering
\includegraphics[width=\textwidth]{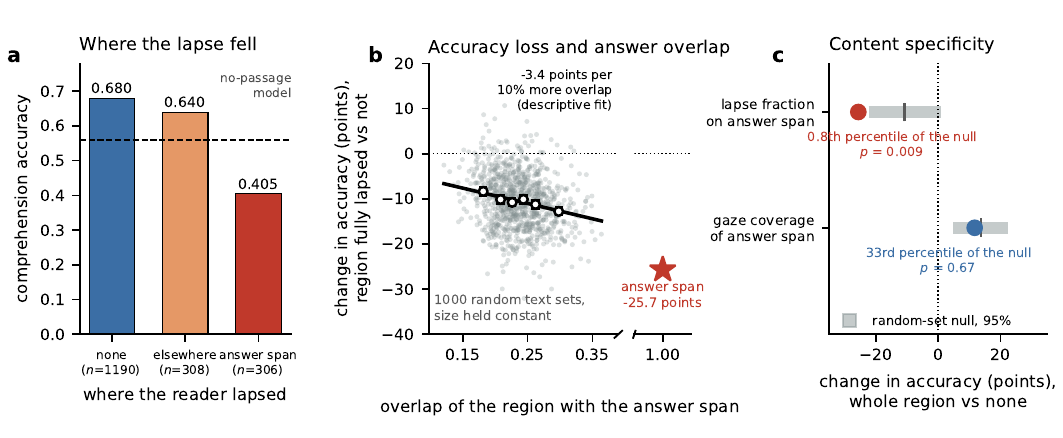}
\caption{\textbf{Reported lapse overlap with answer-bearing content predicts comprehension accuracy.}
Primary sample, 44 readers. (a) Raw accuracy by where the reported lapse fell, with the
no-passage language-model accuracy marked; these are unadjusted proportions and the modeled effect
is smaller. (b) The change in accuracy for a lapse over the whole region, for 1000 sentence-sampled random text sets of the same word count from the same page, against each region's overlap with the
annotated answer span. Because overlap is averaged over items, placements concentrate between 0.12 and 0.36; the axis is broken so that
this sampled range is shown at scale and the true answer span (star) is not read as the end of
an extrapolated line. Open circles are binned means with descriptive 95\% intervals across random draws; the line summarizes the sampled range. Draws reuse the same trials, so these intervals do not measure uncertainty across readers. (c) The same comparison for the report and for gaze coverage (the
fraction of the region's words fixated). Bars are the central 95\% of each random-set null, and ticks are its median. In (b) and (c) effects are percentage points of accuracy for the whole region
lapsed (or fixated) against none of it, from models that also adjust for page coverage (the
report) or for coverage of the fixed non-answer control (gaze).}
\label{fig:comp}
\end{figure}

Despite persistent lexical guidance, reported lapses were associated with poorer comprehension. The dataset includes one multiple-choice question per page. This question provides an outcome that is independent of the physiological recordings (Fig.~\ref{fig:comp}). At page level, accuracy fell from 0.673 to 0.552 on pages with a reported span. With reader and story held
constant, the accuracy loss was 5.9 percentage points ($p=0.006$). The loss grew with the fraction of the page covered ($p=7.2\times10^{-4}$) and was specific to the page whose question is
asked.

We next localized the loss within each page. Each item was annotated with its answer-bearing
sentences, a median of 21\% of the page. An answerability check supported the annotation: a small language
model (Qwen2.5-1.5B-Instruct) given only those sentences chose the keyed answer on 90\% of items,
against 85\% with the whole page and 56\% with no passage (SI~S12).

Our primary estimate uses the fraction of answer-span fixations inside a reported lapse, with reader and item fixed effects and covariates for lapse coverage elsewhere and page coverage. A change from none to all of the answer span inside a lapse was associated with 25.7 percentage points lower accuracy (95\% CI $[16.5, 34.9]$, $p=4.34\times10^{-8}$). This is a model coefficient across trials, not a comparison that every reader experienced. Binary any-lapse coding and additional viewing covariates give related estimates with different meanings; SI~S12 reports them together.

For the content-specificity test, we compared the answer span with 1000 random text sets of the same word count from the same page. Sentences were sampled in random order and the final sentence was truncated to match the required count; the resulting sets need not be contiguous. The median estimated loss was 10.8 points. Only 8 of the 1000 sets matched or exceeded the answer-span loss, giving $p=(8+1)/(1000+1)=0.009$. These sets could overlap the answer span. A separate fixed non-answer control excluded answer-bearing sentences and matched word count in 48 of 50 items; its estimated loss was 0.9 points. Its different content explains why it need not match the random-set median (SI~S12).

Across a separate set of 1000 random draws, the estimated loss was 3.4 points larger for each additional 10\% overlap with the answer span. This gradient describes the resampling distribution; the random-set test above provides the inferential comparison. Adding answer-span gaze coverage, fixation count and dwell time left the lapse association significant (SI~S12).

Gaze coverage does not show answer-specific localization. Gaze coverage of the answer span predicts accuracy in a model without lapse reports (all of the span rather than none: $+11.7$ points, $p=0.016$). However, coverage of the fixed non-answer control predicts accuracy similarly ($+9.7$ points, difference $p=0.77$).
Against the random-set distribution, the true answer span is at the 33rd percentile ($p=0.67$),
and the second annotation yields the same null at the 33rd percentile. Reported lapse overlap with the answer span predicts accuracy; gaze coverage does not show comparable content specificity.

This null is specific to within-page localization. \citet{southwell2020} predict comprehension of long connected
texts from seven global page-level gaze features, and their model replicates on these readers
(median held-out $r=0.338$, against $0.009$ with the outcome shuffled). Within reader, both records contain information about which page had a wrong answer ($r=0.130$ for
the gaze prediction, $r=-0.101$ for a reported lapse, both $p<10^{-5}$). The difference in content specificity appears within the page (SI~S13).

The increase in incorrect answers was not accompanied by a reliable increase in ``not sure'' responses (SI~S12). The effect is large on the trials that it affects but
small in total. Only 17.0\% of trials have a lapse on the answer span. After adjustment for
reader, item, reading amount and mind-wandering elsewhere, the model-based attributable loss was
1.9 points (95\% CI $[1.0, 2.9]$). This is 5.1\% of all errors.

The item bank was designed to encourage on-task reading rather than to isolate comprehension processes. A language model answered 56\% of the questions without seeing the passage. This is evidence that some items are answerable without the text, not a ceiling on human performance. Absolute effect sizes should be interpreted with that limitation; the random-set comparison holds item identity constant.

\subsection{The skipping marker depends on its definition}
\label{sec:skip}

\begin{figure}[!tbp]
\centering
\includegraphics[width=0.9\textwidth]{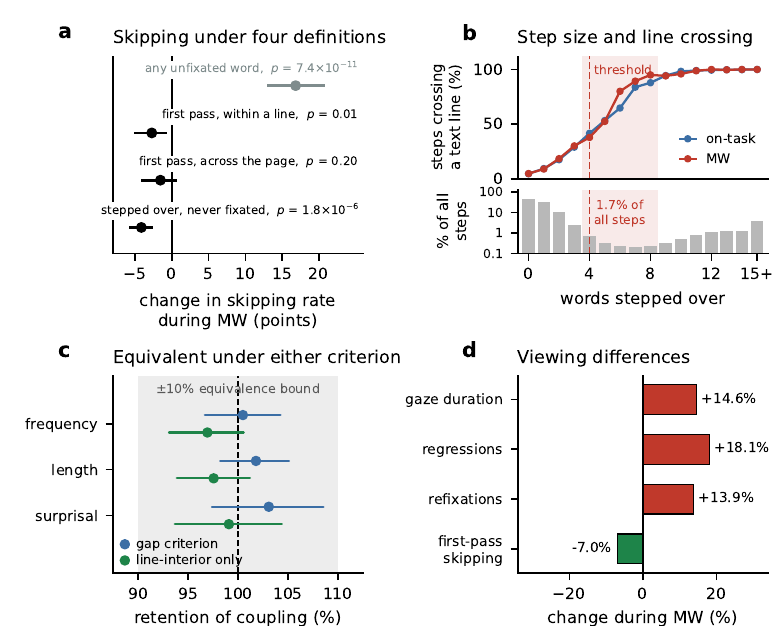}
\caption{\textbf{The skipping marker depends on the definition, and the corrected profile shows
more fixation time, not less.}
Primary sample, 44 readers. (a) Change in the skipping rate during mind-wandering under four
definitions, with 95\% intervals over readers: any word without a first-pass fixation (grey); the
first-pass definition with steps of at most four words, within a text line or across the page; a
word inside such a step that was never fixated.
(b) Top: percentage of forward steps that cross a text line, at each step size. Bottom: how
common each step size is, on-task steps, log scale. The dashed line marks the four-word threshold. The shaded band marks the trough between
within-line reading and return sweeps, which contains 1.7\% of steps. (c) Retention of skipping
coupling under the gap criterion and under a layout-based criterion
using only line-interior words, with 95\% intervals. The grey band is the $\pm$10\% equivalence bound of Section~\ref{sec:preserved}. All six intervals
are inside it.
(d) Mean within-reader change in gaze duration, regressions, refixations and first-pass skipping. Returns to skipped words are omitted because their increase was not reliable ($p=0.13$; SI~S14).
Skipping is first-pass skipping within a line. Threshold sweeps and the off-text time diagnostic
are in SI~S4.}
\label{fig:measurement}
\end{figure}

The most consistent behavioral marker in the meta-analytic literature appears to conflict with
these nulls: readers skip more words when their minds wander \citep{meziere2025}. Our data
reproduce that result under the usual definition in continuous reading, but not when the scan
path has to step over the word (Fig.~\ref{fig:measurement}a).

Scoring a word as skipped whenever it received no first-pass fixation anywhere in the reading span produces a large effect in the expected direction. The skipping rate increases from 0.439 on task to
0.648 during mind-wandering ($\Delta=+0.168$, 95\% CI $[0.131, 0.206]$,
$p=7.4\times10^{-11}$, 41 of 44 readers). The increase disappears when the scan path must step over the word. Under the standard first-pass definition, with steps
of at most four words within a text line, skipping decreased slightly, from 0.372 on task to 0.345 during mind-wandering ($\Delta=-0.027$,
95\% CI $[-0.050, -0.008]$, $p=0.014$). When steps can cross a line break, the decrease was
smaller and not reliable ($\Delta=-0.016$, $p=0.20$). Forward
steps advanced as far in both states (1.66 and 1.64 words per step). When a word counts as skipped only if no fixation ever landed on it, the decrease is larger, from
0.251 to 0.226 ($\Delta=-0.041$, $[-0.057, -0.028]$, $p=1.8\times10^{-6}$, 38 of 44 readers). The
reason is that readers regress more while mind-wandering (below) and thus return to more of the words that they passed over.
Rates are pooled over words and $\Delta$ is the mean of the per-reader differences, which is the
quantity tested. No definition that requires a demonstrated step yields an increase at any
threshold from one to ten words (SI~S4).

The discrepancy comes from words inside large positional gaps, mainly return sweeps across lines and page transitions. A purely layout-based analysis of line-interior words also showed less skipping and preserved coupling; matching exclusion rates across states gave the same conclusion. SI~S4 reports step-size distributions, threshold sweeps and these controls. This is a measurement constraint for continuous reading, not evidence that previous sentence-based studies contain the same artifact.

The corrected profile shows increased viewing during mind-wandering. Relative to
on-task reading by the same readers, gaze durations were 14.6\% longer (95\% CI $[10.3, 19.6]$,
$p=2.2\times10^{-7}$, 40 of 44 readers), regressions were 18.1\% more frequent
($p=1.3\times10^{-4}$, 39 of 44), immediate refixations were 13.9\% more frequent ($p=0.006$, 34
of 44) and first-pass skipping was 7.0\% less frequent ($p=0.014$). Returns to skipped words did not reliably increase; longer lookahead windows introduce unequal opportunity across states (SI~S14). Readers fixated more words for longer, and which
words received that extra time was as strongly coupled to word properties as on task. Tests found no reliable concentration of the additional viewing on difficult words (SI~S8). The 14.6\% value is the mean within-reader increase in word-level gaze duration. It differs from the 5.0\% fixation-level duration contrast and the 10.8\% state coefficient adjusted for reader, token and word properties; these are different outcomes and estimators.

\subsection{Duration coupling is weaker during instructed search}
\label{sec:goal}

Could the duration measure detect reduced lexical guidance? In the independent ZuCo sample, the frequency slope on gaze duration during semantic-relation search was 79\% of its normal-reading value (reader-bootstrap 95\% CI $[67, 91]$; paired $t$-test on the slope change, $p=0.009$), after subtracting the between-session change in sentiment reading. Length coupling fell to 82\% ($p=0.012$); surprisal coupling, at 88\%, did not reliably decrease ($p=0.16$). First-pass skipping was 3.7 percentage points higher during search, with no reliable change in skipping coupling (SI~S10, Fig.~\ref{fig:states}).

Duration coupling differed between these reading tasks. Preserved skipping coupling alone therefore does not establish intact reading. The combined skipping and duration results distinguish the primary-sample lapse pattern from this instructed search pattern. Differences in readers, materials and recording sessions limit a causal comparison between the datasets.

\subsection{Proportional rescaling of the fixation-related response}
\label{sec:state}
\label{sec:bridge}

\begin{figure}[!tbp]
\centering
\includegraphics[width=\textwidth]{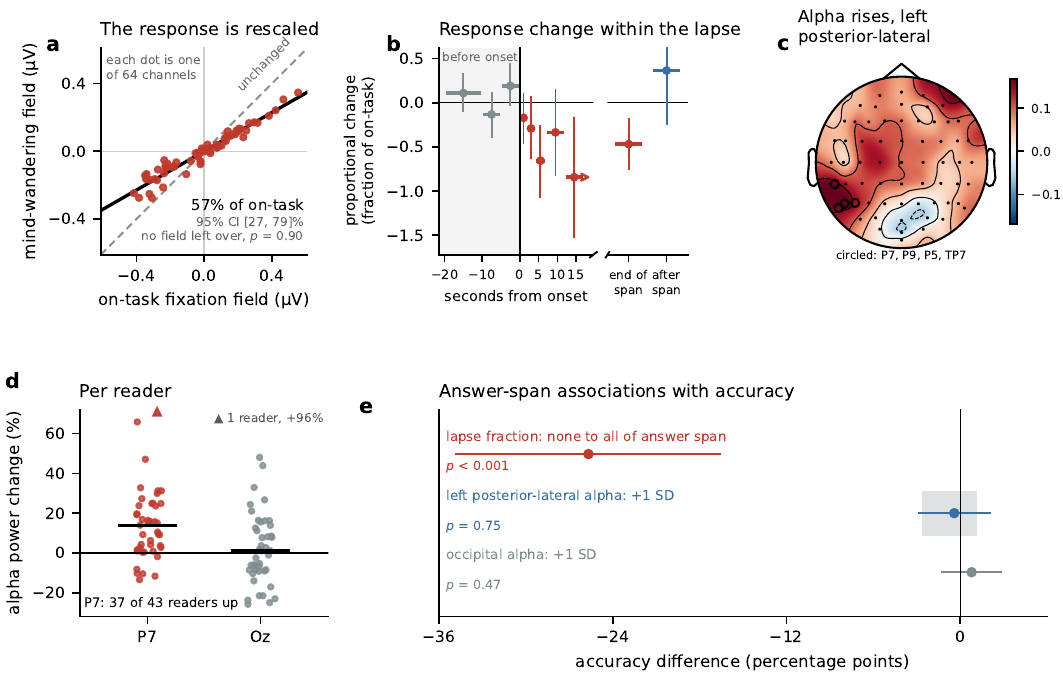}
\caption{\textbf{Neural state changes accompany reported lapses; answer-span alpha does not predict accuracy.}
Primary sample. The fields in (a) are average-referenced and averaged over 150--290\,ms of the deconvolved intercept kernel. (a) The mind-wandering fixation field against the on-task field,
one point per channel. The dashed line is equality. The solid line is the leave-one-reader-out rescaling fit, on which
the mind-wandering field is the on-task field at 57\% of its amplitude. $p$ is the sign-flip test on the field remaining after the rescaling is removed.
(b) Proportional change in the fixation-related response by time from the marked onset, as a
fraction of the on-task field. A value of 0 is no change, and $-0.5$ is a response at half its
size. Bins are unequal in width and drawn at their true extent. Horizontal bars are the averaging
window, vertical bars are 95\% intervals, and the arrow is the open-ended bin beyond 12\,s. The
two points past the break are referenced to the span offset: the last three seconds inside the
span, and on-text reading after it that ended more than 5\,s before the report (SI~S7). (c) Alpha
change during mind-wandering, clean contrast. Circled channels (P7, P9, P5 and TP7) are significant after family-wise
correction over all 64. Map interpolation and markers use the same projected BioSemi coordinates. (d) Alpha change at P7 and Oz per reader before blink/report-period exclusions, medians in black. The axis is capped at $+72$\%; one reader at $+96$\% is drawn as a caret and is kept in
the median and the count. (e) Associations with accuracy in percentage points, with reader-clustered 95\% confidence intervals. The report row shows the primary full-span versus no-span contrast from Section~\ref{sec:comp}. Alpha rows show changes per SD of answer-span alpha, controlling for alpha elsewhere and page coverage. These predictor scales differ and are labeled; their coefficient magnitudes should not be compared directly. The grey band is the random-set null for the left posterior-lateral alpha coefficient.}
\label{fig:globalstate}
\end{figure}

The clearest neural signature of mind-wandering is in the intercept kernel. This kernel is the
deconvolved response common to every fixation, regardless of the word, taken at all 64 channels
(Fig.~\ref{fig:globalstate}a). Over 150 to 290\,ms, the difference field (mind-wandering minus
on-task) is reliable (global field power $0.103$\,\si{\micro\volt}, sign-flip
$p=2.0\times10^{-4}$). It closely resembles a negative scaled copy of the on-task field ($r=-0.944$, reader-bootstrap percentile 95\% CI $[-0.953, -0.662]$; 10,000 draws). This interval is asymmetric because it uses empirical bootstrap percentiles, not a symmetric interval around $r$. A single rescaling factor, fitted leave-one-reader-out, gives an attenuation
of 42.7\% (95\% CI $[21.4, 72.9]$) and accounts for 89.2\% of the variance of the group difference
map. After we subtract it, no reliable residual field remains ($p=0.90$). The measured response is
approximately 43\% smaller, with a scalp distribution well described by proportional scaling. In other words, the mind-wandering
field is the on-task field at 57\% of its size (SI~S7).

The attenuation estimate depends on the deconvolution penalty. SI~S7 reports sensitivity to its strength and scaling; the residual field was not significant under any tested scheme.

Measured differences in the incoming saccades did not explain the rescaling. Mind-wandering
barely changed the saccade mix, and the attenuation was 42.4\% with the incoming saccade as a
covariate (95\% CI $[18.9, 67.9]$) and 49\% on forward saccades matched in amplitude across
states, against 46\% for random subsets of the same size (SI~S7).

The rescaling concerns the response common to fixations, whereas the frequency interaction had a positive but imprecise estimate (Section~\ref{sec:preserved}). The data therefore support an amplitude change without resolving every word-property interaction. The fitted state-level factor is not a direct measure of which content was encoded successfully.

Time-resolved estimates showed attenuation within the reported span, with a reliable reduction from 4 to 7\,s after marked onset ($-0.66$, $p=0.004$). Estimates before onset and after offset were imprecise, and page advances without a lapse showed no comparable change. The data do not establish the latency of attenuation or recovery (Fig.~\ref{fig:globalstate}b; SI~S7).

An encoding-model reanalysis found that an amplitude term accounted for much of the observed ratio of fixation-related brain--language alignment between states. Its decomposition and limitations are reported in SI~S11; it does not independently establish preserved representational content.

\subsection{Alpha is higher over left posterior-lateral sensors}
\label{sec:alpha}

Ongoing alpha provided a second neural correlate of reported lapses (Fig.~\ref{fig:globalstate}c, d). After blink and report-period exclusions, four left posterior-lateral sensors were significant under family-wise correction over all 64 channels: P7, P9, P5 and TP7. At P7, alpha increased by a median of 10.0\%, with increases in 35 of 41 readers. Before these exclusions, the effect extended more widely, with P7 at $+14.0\%$ (37 of 43 readers, family-wise $p=4\times10^{-4}$). Occipital sensor Oz showed no reliable change ($+1.1\%$ before exclusions, $-3.6\%$ after; family-wise $p\ge0.999$).

The alpha effect persisted after subtracting fixation-evoked activity, retaining 90--101\% of its size across variants, with P7 significant after correction ($p\le0.006$). Its scalp distribution and between-reader variation did not reliably follow the response-rescaling estimate (SI~S9). These checks support an ongoing alpha effect that is not explained by the smaller fixation-related response. The posterior-lateral distribution motivates a regional interpretation, considered below; it does not localize a cortical generator or establish that the lateral and occipital effects differ statistically.

\subsection{Answer-span alpha does not predict comprehension accuracy}
\label{sec:mark}

We next asked whether alpha measured on the answer span predicted accuracy (Fig.~\ref{fig:globalstate}e). The rescaling model estimates one factor across fixations within a state; we did not obtain a content-resolved estimate with which to test comprehension.

Alpha can be measured per fixation, so we applied the design of
Section~\ref{sec:comp}. Left posterior-lateral alpha, averaged over the fixations on the answer span, did not predict
accuracy ($-0.4$ percentage points per SD, 95\% CI $[-2.9, +2.1]$, $p=0.75$), while a reported lapse on the
same span did. Alpha added nothing beyond the report ($p=0.83$)
and nothing on trials with no reported lapse on the answer ($p=0.52$). Against the random-set
null of alpha, the answer span was at the 68th percentile. The occipital control was null as well ($p=0.47$).

The average alpha shift was small ($+0.06$\,SD per fixation), so the trial-level signal was weak. This test did not establish answer-specific prediction and cannot rule out a contribution of alpha to the lapse.

\section{Discussion}

Word properties continued to guide eye movements during reported lapses associated with poorer comprehension. Skipping retained its association with frequency, length and surprisal, and lexical influences on gaze duration persisted when the same corpus words were compared across states. Readers nevertheless answered less accurately when the reported lapse overlapped answer-bearing sentences. Word-sensitive eye movements did not ensure successful use of the text.

\paragraph{What continued during a lapse.}
Figure~\ref{fig:design}a contrasts reduced lexical guidance with persistent guidance despite impaired comprehension. The joint skipping and duration results favor persistent guidance for sustained, self-caught lapses, within the tested 10\% and 20\% bounds, respectively. This conclusion is stronger than observing continued eye movements: those movements remained sensitive to the words encountered. It is also narrower than demonstrating intact word recognition or semantic processing. The frequency-related EEG estimate is imprecise, and the surprisal contrast cannot distinguish a preserved from an abolished N400. The results constrain formulations of perceptual decoupling that require weaker lexical influences on eye-movement control; they do not reject every form of perceptual or systems-level decoupling.

The search-task comparison shows why the two eye-movement measures must be considered together. Duration coupling was weaker during semantic-relation search, whereas skipping coupling showed no reliable change. Thus, skipping alone does not distinguish reading states. During primary-sample lapses, both associations persisted even as gaze durations lengthened and regressions increased. Additional viewing was not sufficient for comprehension, and the return measures do not demonstrate successful repair. The skipping audit further shows that an apparent increase in skipped words can depend on how large scan-path gaps are classified in continuous text.

\paragraph{A working hypothesis about comprehension.}
One explanation is that word properties continue to guide the eyes while readers form or retain a less useful representation of the text. This is consistent with earlier work linking mind-wandering to impaired situation-model construction \citep{smallwood2008attention}. It explains why lexical guidance can persist and why the location of a lapse matters for a later content question. It also accommodates increased viewing: spending more time on words need not restore their contribution to a passage representation.

We did not measure the stage at which comprehension failed. The outcome cannot distinguish deficient encoding, integration, maintenance or retrieval. The preserved eye-movement associations establish continued influence of word properties, not successful identification of every word. A follow-up could combine eye tracking with separate tests of immediate word recognition, cross-sentence integration and delayed memory during the same lapse.

\paragraph{Interpreting the alpha distribution.}
The alpha increase was concentrated over left posterior-lateral sensors after blink and report-period exclusions: P7, P9, P5 and TP7. This distribution makes a reading-related interpretation plausible. Visual word processing recruits left occipitotemporal cortex, and MEG work has linked awareness of masked words to alpha suppression in left temporal and parietal regions \citep{gaillard2006reading,magazzini2016alpha}. Increased alpha here could therefore reflect reduced engagement or altered gating within systems supporting the use of written input. Alpha-based inhibition is one proposed mechanism, rather than an interpretation uniquely determined by power \citep{jensen2010gating}. Our scalp measurements cannot identify the generator, demonstrate inhibition of a reading region, or show that the effect is specific to language.

We found no reliable alpha increase at Oz, so this sensor provides no evidence of a general occipital increase during a lapse. This pattern is compatible with the behavioral evidence that visual text continues to influence eye movements while comprehension deteriorates. However, a nonsignificant Oz contrast does not establish unchanged visual-cortical processing, and significance at lateral sensors but not Oz is not a direct test of their difference. The positive evidence for continued word-sensitive eye movements comes from the coupling tests.

Alpha also did not predict accuracy when measured on the answer span. The small trial-level shift limits this null, so it cannot rule out a contribution to the lapse. Similarly, the smaller fixation-related response establishes an amplitude change at the scalp, not a loss of a specific representation. The neural changes accompany continued coupling, but do not explain which answers readers miss. Systems-level decoupling from visual cortex during internally generated thought could coexist with this pattern \citep{zhang2022dmn}.

\paragraph{Scope and limitations.}
The lapses were caught and marked by readers, restricting inference to events that reached meta-awareness. Differences from pre-report studies \citep{reichle2010,schad2012} may reflect which portion or depth of a lapse is sampled. Retrospective span boundaries are uncertain, although label shifts and position-bin analyses left the principal skipping result unchanged. Local skipping eligibility is state-dependent; layout-based and exclusion-matched analyses reduce this concern without eliminating it.

The content result is an association estimated across trials, with one question per page. Automatically annotated answer spans agreed across two procedures but lack human inter-annotator validation. Item answerability without the passage limits interpretation of absolute effect size; the lapse-associated loss accounts for an estimated 5.1\% of all errors. The independent task sample does not establish a causal effect of reading goal relative to mind-wandering. Neural estimates depend on regularization and residual ocular activity, and the semantic contrast remains unresolved.

The results favor an explanation of self-caught lapses in which word-sensitive eye-movement control continues while comprehension is less effective. Explaining the failure requires identifying what happens between encountering those words and using them to answer a question.

\section*{Funding}

This work was supported by funding from the Army Research Laboratory's Human-Guided Intelligent
Systems (HGIS) Program (W911NF-23-2-0067), the Army Research Laboratory's STRONG Program
(W911NF-19-2-0139, W911NF-19-2-0135, W911NF-21-2-0125), the Air Force Office of Scientific
Research (FA9550-22-1-0337), and a Vannevar Bush Faculty Fellowship from the US Department of
Defense (N00014-20-1-2027). The funders played no role in study design, data collection,
analysis and interpretation of data, or the writing of this manuscript.

\section*{CRediT authorship contribution statement}

\textbf{Anirudh Natarajan:} Conceptualization, Methodology, Software, Formal analysis, Data curation, Visualization, Writing -- original draft, Writing -- review \& editing.
\textbf{Paul Sajda:} Supervision, Funding acquisition, Writing -- review \& editing.

\section*{Acknowledgments}

We thank the authors of the ROAMM and ZuCo datasets for collecting and openly releasing the
recordings analyzed here. Those authors are not responsible for the present analyses or conclusions.

\section*{Data availability}

Both datasets analyzed here are publicly available and were collected by others. ROAMM is
distributed by OpenNeuro as \texttt{ds007629} under a CC0 license
(\url{https://openneuro.org/datasets/ds007629/versions/1.3.0}). ZuCo 1.0 is distributed through
the Open Science Framework (\url{https://osf.io/q3zws/}). Stored estimates and figure data are available with the analysis code. They allow the reported summaries to be inspected. Recomputing the estimates requires the inputs described below.

\section*{Code availability}

Analysis code and stored results are available at \url{https://github.com/anirudhmn/MindWandering} (version 2.0.0), under the MIT license. The repository includes the preprocessing and deconvolution pipelines, the analysis scripts for every section of the paper and the supplement, the derived tables, the figure-rendering script and a script that checks the reported numbers against the stored results.

Some results can be recomputed from the derived tables. Others require the public recordings or additional inputs: stimulus word coordinates and comprehension scores for the answer-span analyses, raw recordings for the control-sample potentials, and large intermediate EEG, epoch and language-model arrays for SI~S7, S9 and S11. Those intermediate arrays are not redistributed. Stored result files permit inspection of the reported estimates but are not a substitute for rerunning those analyses. Randomized procedures use fixed seeds recorded in the scripts.

\section*{Ethical statement}

The analyses in this manuscript use two previously collected, publicly available datasets
\citep{roammdata,hollenstein2018}. No new data were collected. The original data collections
were approved by the institutional review boards of the institutions that conducted them, and
written informed consent was obtained from all participants prior to those experiments.

\section*{Conflict of interest statement}

The authors declare that they have no known competing financial interests or personal
relationships that could have appeared to influence the work reported in this paper.

\section*{Declaration of generative AI and AI-assisted technologies in the manuscript preparation process}

During the preparation of this work the authors used Claude (Anthropic) to help write and edit portions of the analysis code and to edit the text of the manuscript. A language model (Claude Opus 5) also marked the answer-bearing sentences of the comprehension items, as described in Section~\ref{sec:methods-comp}. OpenAI Codex was also used to check journal requirements and prepare the submission files. After using these tools, the authors reviewed and edited the content as needed and take full responsibility for the content of the publication.


\end{bibunit}

\clearpage
\setcounter{figure}{0}
\renewcommand{\thefigure}{S\arabic{figure}}
\setcounter{table}{0}
\renewcommand{\thetable}{S\arabic{table}}

\setcounter{page}{1}
\renewcommand{\thepage}{S\arabic{page}}
\suppcitations
\begin{bibunit}[unsrtnat]
\section*{Supplementary Information}
\begin{center}
\textbf{Mind-wandering preserves word-level coupling while disrupting comprehension}\\[0.4em]
Anirudh Natarajan and Paul Sajda
\end{center}

The supplement has its own reference list, numbered from 1. Section and figure references retain the manuscript numbering. The analyses are grouped by the claim or methodological concern they address.
Sections S1 and S2 assess the sensitivity of the measures and the placement of the reported
spans. Sections S3 to S6 test alternative explanations for the preserved-coupling results,
including the skipping diagnostics in S4. Sections S7 to S9 describe neural and behavioral changes during a lapse, summarized in Sections~\ref{sec:state} and~\ref{sec:alpha}. Section S10 concerns the control sample.
Sections S11 to S13 connect the present results to two published analyses and report robustness checks for comprehension. Section S14 presents diagnostics for the return window; S15 reports the full coupling-precision table, and S16 provides extended methods.

\paragraph*{Part I. Baseline associations and reported lapse spans}

 \subsection*{S1. Baseline coupling during on-task reading}

\begin{figure}[!tbp]
\centering
\includegraphics[width=0.78\textwidth]{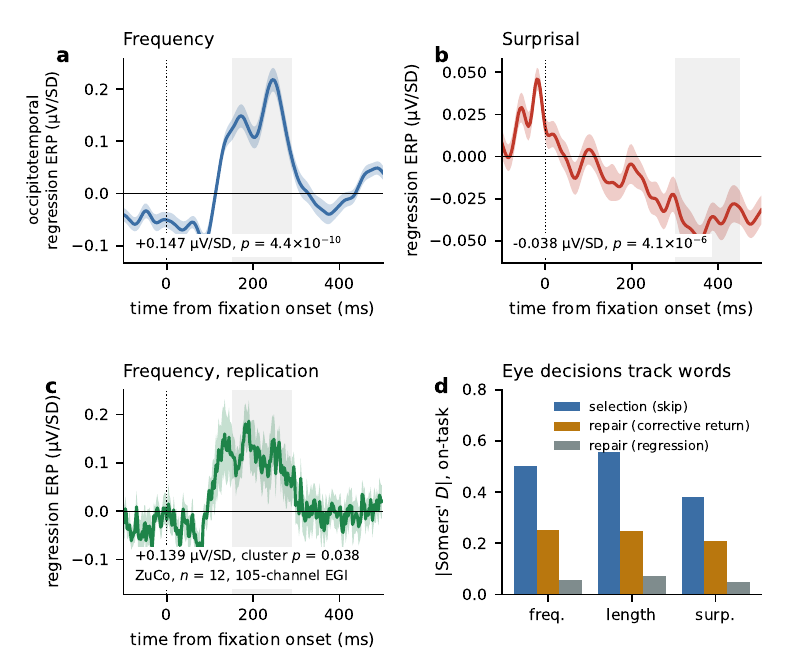}
\caption{\textbf{The measures respond to word properties during on-task reading.}
(a) Overlap-corrected regression-ERP kernel for word frequency at occipitotemporal sensors.
(b) Kernel for contextual surprisal at centroparietal sensors. Shading is the standard error across 44 readers, and gray bands mark the analysis windows. (c) The frequency kernel in
an independent 12-reader sample recorded with different hardware. (d) Rank-based association
between word properties and each of the three eye-movement decisions during on-task reading.}
\label{fig:instrument}
\end{figure}

Before asking whether coupling changes, we established that each measure responds to word
properties while readers are on task (Fig.~\ref{fig:instrument}).

Fixation durations changed with word properties in a mixed model with by-reader random slopes: log
fixation duration decreased with word frequency ($\beta=-0.0178$, SE $=0.0025$,
$p=4.6\times10^{-13}$) and increased with GPT-2 contextual surprisal ($\beta=+0.0159$, SE
$=0.0016$, $p=1.9\times10^{-23}$).

Word properties also predicted the other two decisions. We used Somers' $D$, a rank-based association measure that is insensitive to base rates. Frequent
and short words were more likely to be skipped ($D=+0.50$ for frequency, $D=-0.56$ for length), and surprising words were
less likely to be skipped ($D=-0.38$). Corrective returns to a skipped word were more likely when the skipped
word was rare, long or surprising (frequency $D=-0.25$, $p=3.4\times10^{-11}$; length $D=+0.25$, $p=1.7\times10^{-13}$; surprisal
$D=+0.21$, $p=9.5\times10^{-14}$). Such a pattern is compatible with sensitivity to processing difficulty; it does not establish successful repair.

Fixation-related potentials showed the two canonical effects after overlap correction by
regression ERP \citep{smith2015,ehinger2019}. The deconvolved frequency kernel produced an
occipitotemporal response between 150 and 290\,ms ($+0.147$\,\si{\micro\volt} per SD,
$t_{43}=8.03$, $p=4.4\times10^{-10}$), and the surprisal kernel produced a centroparietal
negativity between 300 and 450\,ms ($-0.038$\,\si{\micro\volt} per SD, $t_{43}=-5.27$,
$p=4.1\times10^{-6}$). The response to word frequency replicated in the independent 12-reader sample, which was recorded
on different hardware in a different laboratory ($+0.139$\,\si{\micro\volt} per SD,
sign-flip cluster $p=0.038$). The surprisal N400 did not replicate there ($p=0.82$; cluster $p=0.22$). The smaller sample and different materials limit this comparison. As a result, only the primary dataset tests the semantic channel.

\subsection*{S2. Validity and placement of the reported spans}

The primary sample uses self-caught reports with reader-marked spans, not probes. If the spans were displaced relative to the state that they mark, a null based on those labels
would be weak. The reason is that misclassification biases any state difference toward zero.

We therefore shifted each reader's mind-wandering indicator by $j$ word positions within run,
for $j$ from $-15$ to $+15$, and recomputed the primary skipping-coupling contrasts at every offset.
The shift preserves the number of mind-wandering words. Sensitivity is thus constant across
offsets, and the comparison is not confounded with power. Retention lay between 96.5\% and 104.4\% for all three word properties at every offset, and every interval spanned zero, so
no offset produced a reliable attenuation. This shows that the null remains under a displacement of up to fifteen words in either direction.
It does not show that the marked spans are not displaced, and it provides no information about a systematic offset larger than that range.

Because the shift keeps the number of mind-wandering words constant, it tests displacement and not
dilution. A short, strongly decoupled interval followed by a long tail would average to the null
that we report. To test this, we cut each span of at least fifteen words into position bins, by
relative third and by absolute distance from each end. In each bin, we recomputed skipping coupling
contrasts against the same reader's on-task value. Retention ranged from 88.5\% to 108.5\% in
every cell. The cells nearest 90\% are in the interiors of long spans, not at either end.
Among the 26 readers with an estimate in both, the first and last thirds differ for no property
(frequency $\Delta D=-0.016$, $p=0.51$; length $+0.014$, $p=0.50$; surprisal $+0.012$,
$p=0.68$). Finally, we pooled over readers with reader fixed effects and reader-clustered standard errors. In
this model, the change in the skipping coupling of a property from the start of a span to its end
was $+4.6\%$ of that coupling for frequency (95\% CI $[-8.1, +17.3]$), $+1.7\%$ for length
$[-8.3, +11.8]$ and $+0.7\%$ for surprisal $[-13.1, +14.6]$. One of the nine paired comparisons
reaches $p=0.057$ uncorrected and is not corroborated by either better-powered test. Decoupling
confined to an interval shorter than ten words, or to spans too short to cut, is not excluded.

We report one observation about the marked onsets, although we cannot interpret it. An interval
longer than 500\,ms followed the fixation step at the start of a reported span in 6.8\% of cases.
This compares with 0.6\% of steps taken within a state, an elevenfold enrichment. The reader marks
the onset word after catching the lapse. We cannot tell whether these long intervals
cause the lapse, occur at the same time, or cause the reader to put the onset where they do. We
draw no inference from them. We note only that they are at the onset and that the contrasts on
each side of it do not differ.

Two further features of the design are relevant to the same objection. The reports are self-caught
and annotated at the moment of catching, not reconstructed at the end of an article. The marked
onset is thus the reader's own judgment of where the lapse began, not a retrospective
estimate over a long interval. And the primary contrasts do not change when we restrict them to
spans the reader marked as fully off-task (S3), the subset for which the label is least
ambiguous.

 \paragraph*{Part II. Precision and robustness of coupling estimates}

 \subsection*{S3. Robustness of the coupling nulls}

\paragraph{Controls.} A relocation control tested whether the observed contrasts are specific
to reported spans. For each reader, we recorded the number and length of their reported spans. We moved each span to
a random on-task position within the same run and recomputed the whole analysis. With 200
relocations, we got a null distribution for each contrast. The observed values were well inside it
($p_{\text{perm}}=0.81$, $0.28$ and $0.31$ for frequency, length and surprisal). We repeated the primary analysis within tertiles of position through a run, with only the spans marked as
fully off-task, and with each article omitted in turn. The skipping-coupling estimates were similar across skip-definition thresholds from one to ten intervening
words. No variant provided reliable evidence of attenuation.

\paragraph{Refixations, regressions and returns.} Refixation coupling held at 96\% to 99\% of its on-task value (35
readers, all $p>0.5$). The tendency of regressions to target difficult words was 74\% for
frequency ($p=0.25$), 73\% for length ($p=0.056$) and 100\% for surprisal ($p=0.99$; 38 readers).
The length effect is the clearest sign of attenuation among the single-property contrasts, and it
is not significant after Holm correction across the three properties. We could estimate
corrective-return coupling in only 16 readers, as it requires enough single-skip events in
both states within a reader. We treat it as uninformative, and not as evidence either way.

\begin{figure}[!tbp]
\centering
\includegraphics[width=\textwidth]{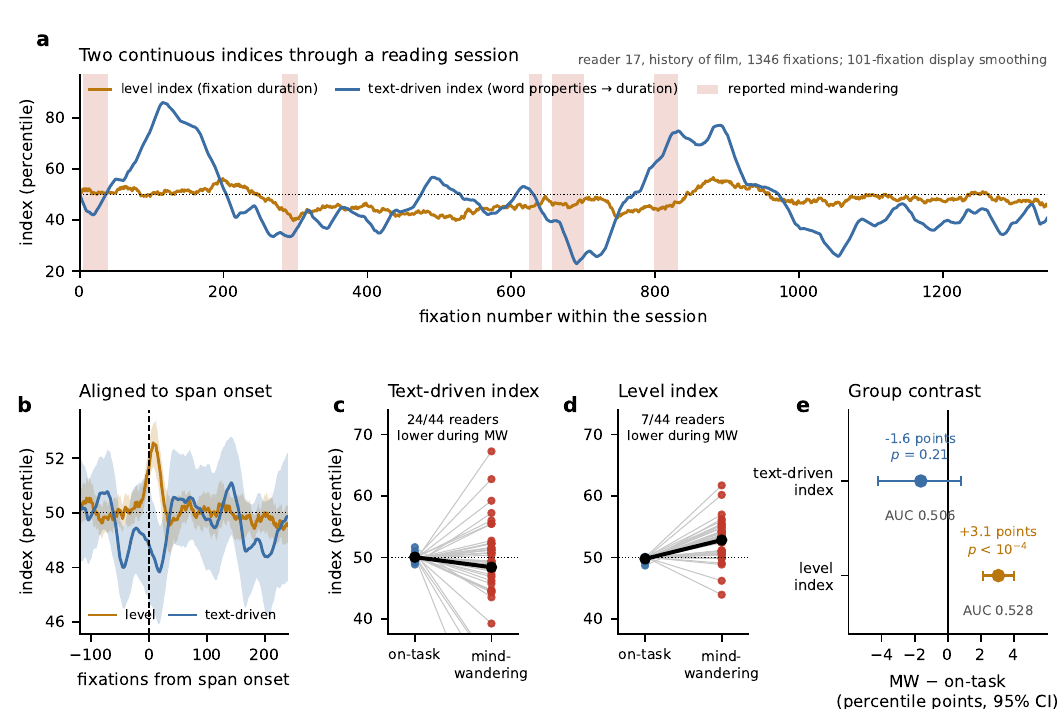}
\caption{\textbf{A continuous coupling index does not reliably distinguish states, but a level index
does.}
Primary sample, 44 readers. (a) Both indices through one reader's session. The text-driven index is the rolling slope of observed on text-predicted log fixation duration,
centered within the window. The level index is the within-reader percentile of raw fixation
duration. Shading marks reported mind-wandering.
Among sessions with at least five reported spans, we show the session whose two state contrasts
are closest to the group means. The traces are smoothed over 101 fixations for display only.
(b) The same two indices aligned to the onset of a reported span, averaged within reader and
then across readers, with 95\% bootstrap bands. This panel is descriptive. The tests are in (c) to (e).
(c, d) Each reader's mean index by state.
(e) The state contrast for both indices, with 95\% bootstrap intervals over readers, and the
area under the curve for classifying individual fixations. The index that never sees a
mind-wandering report is at chance.}
\label{fig:index}
\end{figure}

\paragraph{A continuous index independent of the labels.} Automatic measures of mind-wandering are commonly trained on mind-wandering reports. This makes them unsuitable as independent checks on those reports. We therefore built an index without report labels (Fig.~\ref{fig:index}).

If a reader attends to the text, behavior should stay coupled to word properties. We fitted that
relationship on on-task fixations only. The model predicted log fixation duration from frequency,
length and surprisal with reader effects (Zipf $-0.0128$, length $-0.0014$, surprisal $+0.0028$). Every fixation then has
a duration that the text alone predicts. In a rolling window of 25 fixations we regressed
observed duration on predicted duration, with both centered inside the window. The centering is
what makes the index specific. It removes any overall speeding or slowing. Only one question remains: whether harder and rarer
words currently get proportionally longer fixations.
We scored the result as a within-reader percentile from 0 to 100.

The index did not reliably distinguish mind-wandering from on-task reading. Its mean is 50.1 on task and 48.4 during reported spans, a difference of $-1.6$ points (95\% CI
$[-4.3, +0.7]$, $p=0.18$). The difference was in the predicted direction for 24 of 44 readers. As a classifier
of individual fixations it reaches an area under the curve of 0.506.

The same pipeline applied to a level-based index gives a different result. This index is the
within-reader percentile of raw fixation duration. It increases from 49.8 on task to 52.8 during reported
spans ($+3.1$ points, 95\% CI $[+2.1, +4.0]$, $p<10^{-4}$, area under the curve 0.528).

The contrast isolates two components. The first index measures the slope that relates word
properties to behavior. The second index measures the overall level of behavior. Mind-wandering shifts the
level without a detected slope change. The continuous indices thus agree with the
interaction analyses reported elsewhere in the paper.

\paragraph{Tokenizer mapping.} A word's surprisal is the sum over every sub-word token whose
character span overlaps the word, located by a running cursor through the sentence. A containment rule undercounts words owing to the model's leading-space convention. We note this: the choice is easy to make differently and affects which words have a surprisal value.
As surprisal is conditioned within the sentence, the first word of each sentence has none.

 \paragraph*{Part III. Does the skipping result depend on the exclusion?}

\subsection*{S4. Skipping under alternative definitions}

Section~\ref{sec:skip} shows that the direction of the mind-wandering effect on skipping depends
on how a skip is scored. This section defines each rule, applies all of them to one example scan
path (Table~\ref{tab:skipexample}), and reports the skipping rates under each rule in both datasets
(Table~\ref{tab:skipdefs}). The diagnostics after the tables include two analyses that do not use
the gap criterion.

\paragraph{Definitions.} Every rule starts from the same record: the words of a text in reading
order, and the time-ordered sequence of words on which a reader's fixations landed. A forward step is a move from a fixated word to a later word. The words strictly between are
passed over, and the size of the step is their number. The rules differ in three ways: which words enter the
denominator, whether a word that was passed over and later fixated still counts as skipped, and
which fixation supplies the mind-wandering state of a word that was never fixated. $G$ is the largest step whose passed-over words are scored. We use $G=4$ unless stated.
\begin{enumerate}
\item \emph{Naive.} Every word between the first and the last word the reader fixated in a run is
scored, and it counts as skipped if no fixation ever landed on it. An unfixated word takes the
state of the nearest fixated word in reading order. Nothing requires the scan path to have
passed near the word, so stretches of text that were never read at all, such as the remainder of
a page or a page turned without reading, count as skipped.
\item \emph{Never fixated, no step limit.} The same words and the same skip flags as rule 1, but a
passed-over word takes the state of the fixation that launched the step over it. This rule
exists to separate the effect of the state label from the effect of the gap.
\item \emph{First pass, at most $G$ words.} First-pass progress is the furthest word reached so
far, with words ordered by their first fixation. A fixation that lands beyond it is a first-pass
forward step. If the step passes over at most $G$ words, the word it lands on counts as fixated
and the words it passes over count as skipped, even if the reader fixates them later. Steps larger
than $G$, and words reached only by a regression, are not scored. The step takes the state of the
word it lands on, and steps whose two ends differ in state are dropped. We score it in two ways.
\emph{Across the page}, a step can cross a line break. \emph{Within a line}, each text line is
scored separately, so a return sweep is never a step and the first word fixated on each line is
not scored. This is the standard definition in single-line reading.
\item \emph{Never fixated, at most $G$ words} (the primary analysis). A word counts as skipped if a
step between two consecutive fixations, passing over at most $G$ words, passed over it and no fixation ever landed on it. It counts as
fixated if any fixation landed on it. Words passed over
only by larger steps are dropped. A passed-over word takes the state of the fixation that
launched the step, and the coupling analyses also drop steps whose two ends differ in state
(3.4\% of steps).
\item \emph{Layout.} The words of rule 2, restricted to those lying between the first and the last
fixated word on their own text line. Step size plays no part. What is excluded is the text on a
line before the reader's first fixation on it and after the last, which is where return sweeps
and page turns pass over words. Steps whose two ends differ in state are dropped.
\end{enumerate}
In the control sample each sentence is a single line on its own screen, so line, page and sentence are the same unit. Hence the two versions of rule 3 are the same,
and rules 1 and 5 score the same words. The state is the task condition of the sentence, and rule 2 adds nothing there. The control
data give each word's first-fixation time but not the full fixation sequence, so rule 4 orders
steps by first fixation in that sample.

\paragraph{A worked example.} Table~\ref{tab:skipexample} applies the rules to a three-line passage
and a scan path of twelve fixations. The reader starts on \emph{old}, passes over \emph{the} on
line 1, and steps from \emph{stairs} to \emph{evening} across the line break, passing over
\emph{every}. On line 2 the reader passes over \emph{to} and \emph{the}, reads \emph{great lamp},
and regresses to \emph{the}. The reader then jumps to \emph{the rocks} and never reads the first five words of line 3. Rules 1 and 2 count those five words as skips and give the highest rate,
8 of 20. Rule 3 scores \emph{the} on line 2 as skipped, because the first pass passed over it. Rule 4
scores it as fixated, because the regression later landed on it.

\emph{Every} is
skipped under rule 4 and under rule 3 across the page, but it is not scored within a line or under
rule 5, because it lies before the reader's first fixation on line 2. The rate decreases from 0.40 under rule 1 to 0.14 under rule 5 on the same scan path. The words
that change it most are the words that the reader never came near.

\begin{table}[!ht]
\centering
\caption{\textbf{One scan path scored under each skipping rule.} The passage occupies three text
lines. Fixation order is the position of the fixation in time. S, skipped; F, fixated; --, not
scored. Rules 1 and 2 score the same words with the same flags and differ only in the state label,
which this example does not show.}
\label{tab:skipexample}
\small
\begin{tabular}{rlccccccc}
\toprule
 & & & Fixation & 1, 2 & 3 & 3 & 4 & 5 \\
\# & Word & Line & order & Naive & First pass, page & First pass, line & Never fixated & Layout \\
\midrule
1 & The & 1 & & -- & -- & -- & -- & -- \\
2 & old & 1 & 1 & F & -- & -- & F & F \\
3 & keeper & 1 & 2 & F & F & F & F & F \\
4 & climbed & 1 & 3 & F & F & F & F & F \\
5 & the & 1 & & S & S & S & S & S \\
6 & narrow & 1 & 4 & F & F & F & F & F \\
7 & stairs & 1 & 5 & F & F & F & F & F \\
\addlinespace
8 & every & 2 & & S & S & -- & S & -- \\
9 & evening & 2 & 6 & F & F & -- & F & F \\
10 & to & 2 & & S & S & S & S & S \\
11 & light & 2 & 7 & F & F & F & F & F \\
12 & the & 2 & 10 & F & S & S & F & F \\
13 & great & 2 & 8 & F & F & F & F & F \\
14 & lamp & 2 & 9 & F & F & F & F & F \\
\addlinespace
15 & that & 3 & & S & -- & -- & -- & -- \\
16 & guided & 3 & & S & -- & -- & -- & -- \\
17 & ships & 3 & & S & -- & -- & -- & -- \\
18 & safely & 3 & & S & -- & -- & -- & -- \\
19 & past & 3 & & S & -- & -- & -- & -- \\
20 & the & 3 & 11 & F & -- & -- & F & F \\
21 & rocks. & 3 & 12 & F & F & F & F & F \\
\midrule
 & Skipped / scored & & & 8/20 & 4/13 & 3/11 & 3/15 & 2/14 \\
 & Skipping rate & & & 0.40 & 0.31 & 0.27 & 0.20 & 0.14 \\
\bottomrule
\end{tabular}
\end{table}

\begin{table}[!ht]
\centering
\caption{\textbf{Skipping rates under each rule in both datasets.} $G=4$ throughout. Rates are means over readers of each reader's rate, so each $\Delta$ is the difference between the
two rates beside it. Section~\ref{sec:skip} quotes some rates pooled over words instead, which
weight readers by the number of words they read. Words scored counts word events summed over readers and
states (conditions). $\Delta$ is the mean of the per-reader differences, with a 95\% bootstrap
interval over readers and a paired $t$-test. (a) Primary sample, 44 readers. MW, mind-wandering. The last column counts readers whose rate was
higher during mind-wandering. (b) Control sample, 12 readers.
NR, normal reading (session 1); SR1 and SR2, the two halves of the sentiment task (sessions 1 and
2); TSR, relation search (session 2). The task effect is (TSR $-$ NR) $-$ (SR2 $-$ SR1), which removes session and materials (SI~S10).
The last column counts readers with a positive task effect.}
\label{tab:skipdefs}
\footnotesize
\setlength{\tabcolsep}{4pt}
\textbf{(a) Primary sample: on task versus mind-wandering}\\[2pt]
\begin{tabular}{llrcccrc}
\toprule
Rule & & Words scored & On task & MW & $\Delta$ [95\% CI] & $p$ & Higher \\
\midrule
1 & Naive & 463{,}173 & 0.444 & 0.613 & $+0.168$ [$+0.131$, $+0.206$] & $7.4\times10^{-11}$ & 41/44 \\
2 & Never fixated, no limit & 463{,}173 & 0.469 & 0.338 & $-0.131$ [$-0.161$, $-0.102$] & $3.8\times10^{-11}$ & 5/44 \\
3 & First pass, page & 210{,}788 & 0.399 & 0.384 & $-0.016$ [$-0.041$, $+0.006$] & $0.20$ & 23/44 \\
3 & First pass, line & 262{,}357 & 0.372 & 0.345 & $-0.027$ [$-0.049$, $-0.009$] & $0.014$ & 17/44 \\
4 & Never fixated, $\le G$ & 332{,}467 & 0.250 & 0.209 & $-0.041$ [$-0.057$, $-0.027$] & $1.8\times10^{-6}$ & 6/44 \\
5 & Layout & 340{,}416 & 0.268 & 0.233 & $-0.035$ [$-0.049$, $-0.022$] & $1.1\times10^{-5}$ & 11/44 \\
\bottomrule
\end{tabular}

\vspace{8pt}
\textbf{(b) Control sample: normal reading versus the shallow search goal}\\[2pt]
\begin{tabular}{llrccccccc}
\toprule
Rule & & Words scored & NR & SR1 & SR2 & TSR & Task effect [95\% CI] & $p$ & Higher \\
\midrule
1, 5 & Naive (= layout) & 213{,}121 & 0.289 & 0.253 & 0.290 & 0.394 & $+0.068$ [$+0.047$, $+0.094$] & $2.4\times10^{-4}$ & 12/12 \\
3 & First pass & 169{,}352 & 0.393 & 0.368 & 0.387 & 0.450 & $+0.037$ [$+0.025$, $+0.053$] & $4.6\times10^{-4}$ & 12/12 \\
4 & Never fixated, $\le G$ & 199{,}088 & 0.251 & 0.230 & 0.266 & 0.309 & $+0.023$ [$+0.012$, $+0.034$] & $0.0033$ & 10/12 \\
\bottomrule
\end{tabular}
\end{table}

\paragraph{What the definitions change.} In the primary sample the sign of the effect depends on
the rule, and only the naive rule gives an increase. Rules 1 and 2 score the same 463{,}173 words with identical skip flags. They disagree only about
the state of 24{,}325 words. Of these, 23{,}153 are labeled mind-wandering by the nearest fixated
word but on task by the fixation that launched the step over them, and 1{,}172 are labeled the
other way. Of the 23{,}153, 97.1\% were never fixated and 95.4\% are inside steps of more than
four words, with a median step of 179 words. They are long stretches of unread text whose nearest
fixated word happens to fall inside a lapse. When these stretches get the state of the fixation
that launched the jump, the sign reverses. The naive
increase comes from unread stretches and the state they are given. Every rule that
scores only words the scan path passed over within a few words produces a decrease or no reliable
change.

In the control sample, no text goes unread at a line break or page turn. There, the rules agree in
sign: the shallow search goal increased skipping under all three, and the naive rule gave the
largest task effect. Absolute rates differ by up to 0.2 between rules within a dataset. Skipping
rates from different studies are comparable only under the same rule. Under rule 3 the
on-task rates of the two datasets are close (0.372 within a line in the primary sample, 0.393 for
normal reading in the control sample; $p=0.47$).

\paragraph{Thresholds.} Under rule 4 skipping decreases during mind-wandering at every
$G\in\{1,2,3,4,6,10\}$ ($\Delta$ from $-0.026$ to $-0.043$, all $p<0.001$), and removing the limit
altogether (rule 2) still gives a decrease.

\paragraph{The first-pass definition.} Within a text line, skipping decreased during
mind-wandering at every $G$ ($\Delta$ from $-0.021$ to $-0.028$, all $p\le0.035$). Across the
page, a step of four words or fewer can cross a line break. There, the decrease was reliable only
at $G=1$ ($-0.023$, $p=0.011$; $G=4$: $-0.016$, $p=0.20$). Forward steps within a line advanced 1.66 words on task
and 1.64 during mind-wandering, with the same distribution of step sizes. The larger decrease under rule 4 comes from words that were skipped and later fixated. Readers regress more during mind-wandering, leaving fewer of the words that they pass over stay unfixated.
Skipping coupling under the first-pass definition was preserved within a line (retention 98.4\% $[93.9, 102.6]$ for
frequency, 101.8\% $[98.0, 105.5]$ for length and 99.6\% $[93.0, 106.7]$ for surprisal; TOST at
$\pm$10\% $p=4\times10^{-4}$, $1\times10^{-4}$ and $0.006$). Across the page it was 92.7\%
$[87.8, 97.4]$ for frequency ($p=0.006$), 96.2\% for length and 95.8\% for surprisal
($p\ge0.10$). The page-level frequency estimate is the only skip measure that shows attenuation. It includes the
words passed over at a line break, which the other definitions exclude by design.

\paragraph{Composition of the large gaps.} Steps spanning more than four words cross a text-line
boundary in 96.6\% of on-task cases and 97.7\% of mind-wandering cases, against 4.5\% and 4.7\%
of steps between adjacent words. That contrast is a property of the bins. At each separate step size (Fig.~\ref{fig:measurement}b),
the crossing probability increases smoothly. It is 4.5, 9.2, 17.4, 28.9, 41.5, 53.2, 64.7, 83.8
and 87.8\% for steps of zero through eight words on task, and 98.2\% at ten. The bins look like a
step because the large gaps dominate the $>4$ bin. Of 237{,}943 on-task forward steps, 89.2\% span at most three words and cross a line 8.3\% of the
time, 9.1\% span nine or more and cross a line 99.5\% of the time, and 1.7\% are between. The
mind-wandering proportions are 88.3, 10.3 and 1.4\%. Because crossing rises smoothly, the
four-word threshold is not a natural boundary; it is safe because the definitions disagree only
about the 1.7\% of steps in the trough. We clustered the tops of the word bounding boxes into text lines within each page. The tolerance
was 40 pixels and the line spacing was about 90 pixels, as the tops vary by a few pixels with
glyph height.

\paragraph{Off-text gaze.} Only fixations that the dataset providers could map to a word are
present in the released derivative. We can measure the interval between successive
mapped fixations, but we cannot observe a fixation that fell outside every word bounding box.
Our claim about off-text gaze is an inference from elapsed time. A fixation that we cannot see
would still use time, and the intervals that we observe are too short to contain one. Within a
page, the fraction of time not spent on a mapped word was the same in the two states (0.166 versus
0.169, $p=0.49$). The interval between successive fixations was indistinguishable across states
within every gap size (medians of 26 to 43\,ms throughout).

\paragraph{Composition of the gaps.} During mind-wandering, 74\% of stepped-over words are inside large same-page forward jumps,
against 48\% on task. On task, 29\% of stepped-over words come from page transitions, against
under 1\% during mind-wandering. Reported spans rarely extend across a page change. The naive
measure consequently builds the two states from different kinds of gap.

\paragraph{Differences among excluded words.} Relative to on-task, the words
excluded during mind-wandering were slightly more frequent ($\Delta=+0.16$ Zipf,
$p=1.7\times10^{-4}$), shorter ($\Delta=-0.29$ characters, $p=1.7\times10^{-6}$) and less
surprising ($\Delta=-0.35$ bits, $p=0.023$). This is why we repeated the primary analysis in
two ways that do not rely on the gap criterion.

\paragraph{A purely layout-based rule.} A word is analyzable if the reader fixated at least one word on the same text line both before and
after it. Gap size plays no part. Skipping still decreased during mind-wandering (0.268 to 0.252,
$\Delta=-0.035$, $p=1.1\times10^{-5}$), and skipping coupling was still preserved (retention 96.9\%, 97.6\% and 99.1\%; all $p>0.11$).

\paragraph{Matching the exclusion rate.} When we matched the exclusion rate across states within reader, the coupling estimates were
similar (99.4\%, 98.6\% and 101.0\%; all $p>0.4$). That procedure distorts the composition of the retained set, and we do not interpret a rate contrast from it.

\paragraph{Additional scan-path diagnostics.}
The two definitions differ over words inside large positional gaps in the scan path. Step sizes
are bimodal (Fig.~\ref{fig:measurement}b): 89.2\% of forward steps span at most three words, 9.1\%
span nine or more and almost always cross a line, and only 1.7\% fall between. The words inside
the large steps are line ends and line starts passed over by return sweeps and page transitions,
which multi-line reading studies conventionally exclude \citep{slattery2019}. Gaze was not off the
text during those gaps (SI~S4).

The result does not depend on this exclusion. Under a purely layout-based rule, in which a
word is analyzable if the reader fixated a word on the same line both before and after it,
skipping still decreased during mind-wandering (0.268 to 0.252, $\Delta=-0.035$,
$p=1.1\times10^{-5}$). Skipping coupling was still preserved (retention 96.9\%, 97.6\% and 99.1\%;
all $p>0.11$; Fig.~\ref{fig:measurement}c). When we matched the exclusion rate across states
within reader, the coupling estimates were the same (SI~S4).

\subsection*{S5. The omnibus model-based coupling test}

Each contrast in the main text fixes a decision and a word property in advance. This leaves open the possibility that we chose the measures that do not change. An omnibus model removes this choice. Its
full specification and results follow.

\paragraph{Design.} For every fixation, a network predicted which of the 41 words within twenty
positions the reader would move to next, and how long the current fixation would last. It used two
separate blocks of information: page geometry and reading history on one side, and the word
properties of the candidates on the other. Networks of this kind now predict reading scan paths
well \citep{deng2023eyettention,bolliger2025scandl} and have been used to model task effects
\citep{hahn2023task}. Here, the network is only a measurement instrument. We fitted it on on-task
fixations alone and evaluated it in cells where both the reader and the article were held out. The
measure is how much the word-property block improves held-out prediction, in bits. Another
approach fits a mechanistic model of eye-movement control to an instructed mindless condition and
reads decoupling from a parameter \citep{nuthmann2009}. Our instrument is different: it assumes no
mechanism, and it returns a bound.

The unit is one fixation-to-fixation transition with both words mapped inside the actual reading
interval of the page (392,766 transitions). We dropped the 0.9\% whose target falls outside the
candidate window. The candidate set for each transition is the words at page
positions $\pm20$ from the currently fixated word, which contains the true target on 99.1\% of
transitions.

We fit two heads. The target head scores every candidate and trains with a softmax over the set. The duration head
predicts log fixation duration. Each head receives a geometry block and,
in the compared model, a text block. The geometry block holds page progress, line, position in
line, distance to the line end, gaze position and its offset from the word center, incoming
saccade amplitude, previous fixation duration, first-pass status, ordinal position in the run,
and the number of times this reader has already fixated the candidate on this page. The text
block holds Zipf frequency, length, within-sentence surprisal and the two extended-context
terms, taken at the candidate and at the currently fixated word for the target head, and at the
fixated word and $w-1$, $w+1$, $w+2$ for the duration head.

We trained both heads on on-task transitions only. We evaluated them in a $5\times4$ design of
five article folds crossed with four reader groups. Each of the 20 fitted models is tested on transitions from an article and from readers that it never saw. We compare the target
head as two separately trained models. Its measure per transition is $\log_2 p_{\text{text}} -
\log_2 p_{\text{no text}}$ on the word actually fixated. The duration head has two stages. We fit
a geometry network first and then fit the text block to its held-out residual. The reason is that
the lexical contribution to duration is about 0.2\% of the variance, and the seed-to-seed
variability of two independently trained networks is larger. When we replaced the ridge text stage
with a network, the duration measure became three times smaller (24 of 44 readers positive against
41). We therefore treat the text-to-duration mapping as linear.

Inference is at the reader level. We used per-reader means in each state, a 10,000-draw bootstrap
over readers for intervals, and the same fixed-effect ladder and span-preserving permutation used
elsewhere in this paper. The permutation replaces each reader's mind-wandering labels with the
labels of another reader, transferred by article, page and word position. It keeps span length,
contiguity and the position of spans in the text. It breaks only the link between a reader's own
lapses and their own eye movements.

 \paragraph{Result.} On task, knowledge of the words was worth $0.144$ bits per movement (all 44 readers positive,
$p=1.8\times10^{-25}$). The geometry-only baseline was $2.70$ bits out of the $5.36$ available
from the candidate set. The probability that the model gives to the word actually fixated next
increases by about a tenth. In a further check, we permuted which word is at which position within a
page, with layout, timing and the reader's own movements unchanged. This permutation removes the
improvement completely ($-0.013$ bits, 4 of 44 readers). What the model uses is word identity, not geometry under another name.

During mind-wandering, the measure retained $0.911$ of its on-task value (95\% CI $[0.822,
0.995]$), with a change of $12\%$ detectable at 80\% power. The duration head retained $1.071$
$[0.728, 1.424]$ and is much less informative (Table~\ref{tab:equiv}). The $9\%$ shortfall in
target choice is the most precise attenuation in this study. We do not present it as established: it is not significant under a span-preserving permutation. In
this permutation, we replace
each reader's spans with another reader's spans at the same positions in the same text ($z=-1.84$,
$p=0.067$, 2000 draws). The shortfall remains under the tested geometry controls; we found no reliable difference between movement types.

\paragraph{Robustness.} The per-transition measure is heavy-tailed, with a standard deviation of
$0.68$ bits against a mean of $0.138$ and a median above the mean. Retention is a ratio of means.
For this reason, we re-derived retention under the ten variants below: seven exclusions and three robust
estimators.
Nothing is refitted; every variant reads the same stored per-transition values.

\begin{center}
\small
\begin{tabular}{lrrr}
\toprule
Variant & Transitions & Retention & 95\% CI \\
\midrule
All transitions & 392,766 & 0.911 & $[0.822, 0.995]$ \\
Trimmed outside the 1st and 99th percentile & 384,862 & 0.897 & $[0.812, 0.976]$ \\
Trimmed outside the 5th and 95th percentile & 353,448 & 0.926 & $[0.852, 0.995]$ \\
Winsorized at the 1st and 99th percentile & 392,766 & 0.908 & $[0.822, 0.989]$ \\
Winsorized at the 5th and 95th percentile & 392,766 & 0.909 & $[0.828, 0.985]$ \\
Median, not mean per reader & 392,766 & 0.942 & $[0.878, 1.005]$ \\
Saccades of ten words or fewer & 355,495 & 0.881 & $[0.784, 0.972]$ \\
Saccades of five words or fewer & 347,252 & 0.911 & $[0.812, 1.005]$ \\
Fixations of 80 to 800\,\si{\milli\second} & 382,711 & 0.902 & $[0.809, 0.990]$ \\
Span interiors only & 384,421 & 0.903 & $[0.812, 0.993]$ \\
Forward saccades only & 249,929 & 0.965 & $[0.875, 1.057]$ \\
\bottomrule
\end{tabular}
\end{center}

Retention always stays between $0.88$ and $0.97$. The most outlier-resistant estimator, the per-reader median, yields the smallest shortfall at $0.942$ $[0.878, 1.005]$. The tails contribute to the shortfall but do not produce it.

The shortfall persists under a ladder of fixed effects that absorbs launch line, word identity
and reader by page. Across this ladder, $\beta$ changes only from $-0.015$ to $-0.013$ ($p=0.019$
at the strictest step). The same ladder recovers the known additive lengthening of fixations
during mind-wandering ($+4.2\%$ decreasing to $+2.9\%$, all $p<10^{-8}$). This is its positive
control.

Retention looks higher for forward saccades ($0.965$) than for regressions and refixations
($0.899$). That difference lies in the denominator, not in the effect. In bits, the shortfall is
$-0.0121$ for forward saccades and $-0.0110$ for the other movements, against on-task baselines of
$0.163$ and $0.108$. The within-reader difference of differences is $+0.0012$ $[-0.026, +0.029]$,
$p=0.94$, with 20 of 40 readers in each direction. Forward saccades simply have more text-driven
signal to lose. The movement-type contrast did not distinguish the shortfall between these two classes.

 \subsection*{S6. The neural contrasts in detail}

\paragraph{Cluster tests.} A cluster-based permutation test over the whole epoch and all
sensors returned no cluster for either the frequency-by-state or the surprisal-by-state
interaction ($\min p=1.0$). The cluster tests do not establish equivalence for either kernel.

\paragraph{Power.} The smallest change the surprisal contrast could have detected at 80\% power
is 283\% of the on-task effect, against 118\% for the frequency contrast. The limit is the signal-to-noise ratio of the response to surprisal itself. This response is
reliable at the group level ($p=4.1\times10^{-6}$) but small in absolute terms
($-0.038$\,\si{\micro\volt} per SD). A 30\% modulation of it is far below what 44
readers can resolve. A single ridge penalty can shrink the sparse interaction kernels disproportionately. SI~S7 compares it with the column-scaled penalty used here.

\paragraph{The additive shift, and data quality.} The additive pattern seen in behavior
appeared in the neural data as well: an occipitotemporal amplitude offset of
$+0.136$\,\si{\micro\volt} (95\% CI $[0.076, 0.203]$, $p=1.7\times10^{-4}$) with the slope
not reliably different between states. S7 evaluates the relationship between the state and fixation fields over the whole scalp. With matched trial counts, the split-half reliability of the fixation-related response was
identical in the two states (0.888 versus 0.889, $p=0.91$). This check did not find lower split-half reliability during mind-wandering.

\paragraph{Pooled modeling of the response to surprisal.} We asked whether the uninformative surprisal contrast comes from the analysis and not from the
data. We refitted the fixation kernels so that the text kernel pools across readers. Every
evaluated cell is held out on both the reader and the article. The measure became the improvement
in held-out prediction of the continuous recording over the whole $0$ to $500$\,\si{\milli\second}
window at all 64 sensors. It was no longer a coefficient in one window at one region.

The fitted model passed its validation checks. It predicted held-out data occipitotemporally ($p=1.6\times10^{-7}$, 36 of 44 readers) and
centroparietally ($p=6.3\times10^{-7}$, 37 of 44). When we permuted words within a page, the same quantity fell below zero in two independent draws. The mind-wandering kernel recovered the additive occipitotemporal offset with the same sign
($+0.070$\,\si{\micro\volt}, $p=7.8\times10^{-4}$). That kernel is smaller than the $+0.136$\,\si{\micro\volt} above. The reason is that this model
uses a single ridge penalty across its columns, which shrinks the sparse mind-wandering column
(S7).

This model improved sensitivity, but not enough to resolve the target contrasts. The smallest
detectable change is 82\% of the on-task effect
for the frequency channel, against 118\% for the single-window contrast, and 129\% for the
surprisal channel, against 283\%. Neither is small enough to bound the attenuations that the decoupling account predicts. The state
contrasts from this model are null for both channels ($p=0.35$ and $p=0.29$). A nonlinear mapping
from word properties to the kernel did not improve prediction
($+0.0006$\,\si{\micro\volt\squared}, $p=0.47$, 24 of 44 readers). This agrees with the behavioral
result that a nonlinear text stage is worse than a linear one.

Replacing the word properties with GPT-2 layer representations did not improve sensitivity. The best layer, the fourth, gives a held-out improvement of $0.0068$\,\si{\micro\volt\squared},
against $0.0069$ for the four word properties. Prediction decreases with depth, and the final
layer is below the word-permutation control. No layer could detect a change smaller than 1.9 times
the on-task effect, so the depth profile cannot answer the question.

This analysis raises one caution that applies beyond it. When the text block contains
only word length, the held-out improvement is larger at right lateral frontal sensors
($0.0175$\,\si{\micro\volt\squared}) than at occipitotemporal sensors ($0.0075$). This pattern is
the signature of residual ocular contamination, as length affects saccade amplitude. When we
remove length, the occipitotemporal sensors are again at the top. Any predictor correlated with
saccade amplitude can produce apparent coupling that remains after amplitude nuisance regressors. S7 models
the incoming saccade directly and shows that the state effects do not depend on it.

\paragraph*{Part IV. Neural and behavioral differences during reported lapses}

\subsection*{S7. Proportional rescaling of the measured fixation-related response}

 The amplitude of the fixation-related response differs between states, whereas the word-level coupling measures are preserved. We asked whether the response changes in spatial or temporal
pattern, or keeps its pattern at a smaller amplitude. A change in pattern would suggest a
different configuration of underlying sources.

\begin{figure}[!tbp]
\centering
\includegraphics[width=\textwidth]{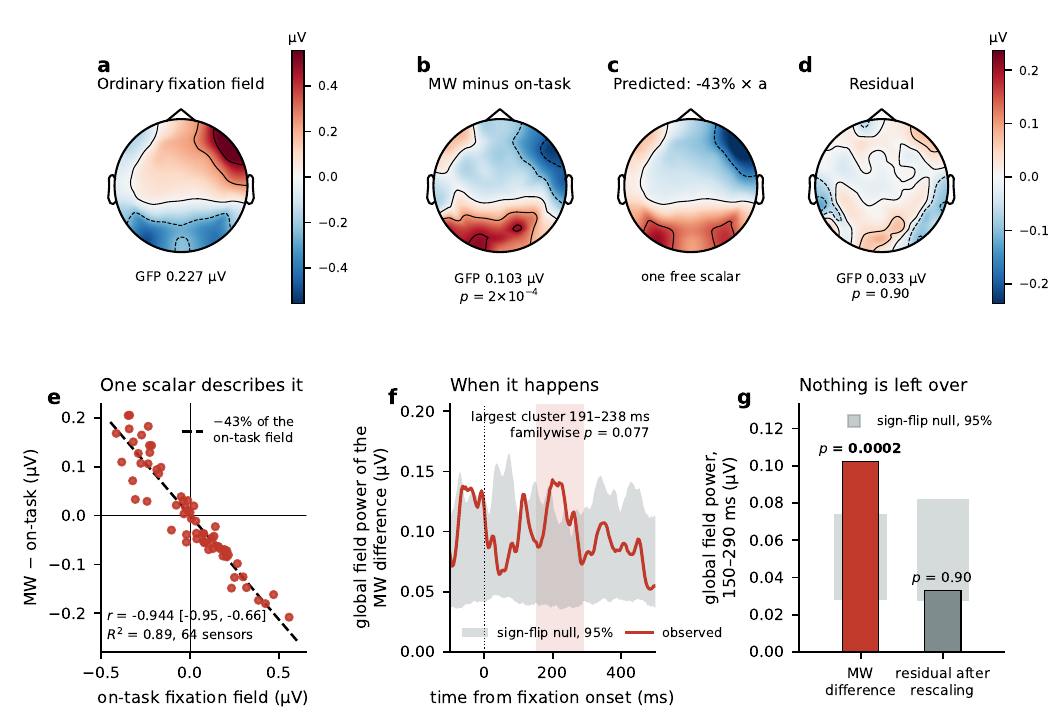}
\caption{\textbf{Mind-wandering rescales the fixation-related response and does not reconfigure
it.}
(a) The overlap-corrected response to a fixated word on task, averaged over 150 to 290\,ms,
as a scalp field.
(b) The difference field (mind-wandering minus on-task) over the same window, on its own color
scale.
(c) What a single proportional scaling of panel a predicts, using the leave-one-reader-out
factor of $-42.7$\%. One free parameter.
(d) The residual after subtracting that prediction.
(e) The same relationship channel by channel. Each point is one of the 64 sensors. The dashed line
is the fitted proportional model.
(f) Global field power of the difference field through the epoch, against a sign-flip null over
readers. Shading marks the analysis window. The largest cluster is not significant after family-wise
correction ($p=0.077$).
(g) Global field power in the analysis window before and after the rescaling, each against its
own sign-flip null. Primary sample, 44 readers. Panels are sensor-level, and we do not interpret them as source localization.}
\label{fig:gain}
\end{figure}

The two possibilities predict different difference maps. If the pattern changes, the difference
between states should have a shape of its own. If only the size changes, the difference should
look like a scaled copy of the ordinary response (Fig.~\ref{fig:gain}).

The difference looks like a scaled copy (Section~\ref{sec:state} gives the main statistics). The
leave-one-out factors range from 36.4 to 47.2\%, and the residual field after rescaling is
$0.033$\,\si{\micro\volt} ($p=0.90$). In the time-resolved analysis, the largest cluster of the difference field extends from
191 to 238\,ms, within the analysis window. It is not, however, significant after family-wise correction
across the five kernels tested together ($p=0.077$). The result depends on the window test above,
and we do not claim a latency for the change.

A single estimate summarizes the observed neural effect on the fixation-related response. Its
scalp pattern stays the same, and its amplitude decreases by 43\%.

\paragraph{Why this is invisible in the raw recording.} Although a 43\% amplitude loss is large, it is absent from ordinary band power (SI~S9). Variance decomposition explains why. The deconvolved
response to all fixations accounts for 3.5\% of the variance of the continuous recording. Power is
proportional to amplitude squared, so scaling that component by 0.57 removes at most 2.4\% of
total power. Because only the early part of the kernel changes, the expected broadband change is closer
to 1\%. That is smaller than the state differences that we can resolve in band power. It is the
same size as the small theta decrease that we observe. A proportional rescaling of the response is
easy to see in a fixation-locked model that averages thousands of events. It is almost invisible
in the spectrum of the raw signal.

\paragraph{Dependence on the ridge penalty.} The deconvolution is regularized, and the state kernels are the most sensitive to the method of
regularization. The mind-wandering columns of the design are nonzero on 7\% of fixations. A single
penalty sized to the average column penalizes them several times more in relative terms than the dense columns. This biases the state
kernels toward zero. Penalizing each column in proportion to its own
energy removes the asymmetry. The table lists every state estimate under both schemes.

\begin{center}
\small
\begin{tabular}{lrrrrr}
\toprule
Penalty & Attenuation & 95\% CI & Map $R^2$ & Residual $p$ & Frequency $\times$ state \\
\midrule
single, $10^{-2}$      & 25.8\% & $[13.8, 39.0]$ & 0.761 & 0.28 & $+16.0\%$ \\
per column, $10^{-4}$  & 43.5\% & $[22.3, 73.5]$ & 0.900 & 0.91 & $+55.2\%$ \\
per column, $10^{-3}$  & 43.5\% & $[22.3, 73.3]$ & 0.899 & 0.91 & $+55.3\%$ \\
per column, $10^{-2}$  & 42.7\% & $[21.4, 72.9]$ & 0.892 & 0.90 & $+56.5\%$ \\
per column, $10^{-1}$  & 36.6\% & $[14.4, 68.4]$ & 0.818 & 0.85 & $+65.1\%$ \\
\bottomrule
\end{tabular}
\end{center}

\noindent
The asymmetry is visible in the maps themselves. Measured against the lightest per-column
setting, the single penalty leaves the on-task fixation field at 98.0\% of its size and the
mind-wandering map at 74.1\%. A per-column penalty of the same nominal strength treats them
alike, at 97.1\% and 97.3\%. The estimator was therefore biased toward finding no change of state. The bias was largest in the
readers who reported the fewest lapses, whose state columns have the least energy.

The per-column estimate is stable from $10^{-4}$ to $10^{-2}$ and starts to shrink only at
$10^{-1}$. We report $10^{-2}$. The $10^{-2}$ per-column row uses the main analysis's bootstrap interval. Other rows use separate resampling draws. The residual field is null under every scheme. No tested penalty left a significant residual field after proportional scaling. The size of the rescaling does depend on
the penalty, and the single-penalty value understates it.

\paragraph{Saccades and ocular activity.} The early fixation-related response increases with the
size of the incoming saccade. A change in the saccade mix could, in principle, look like a smaller
response. We
took each fixation's incoming saccade from the eye-tracker events and classified it from the
fixated words. The mix changed little during mind-wandering (within reader, 44 readers):

\begin{center}
\small
\begin{tabular}{lrrrr}
\toprule
Incoming saccade & On task & Mind-wandering & $\Delta$ & $p$ \\
\midrule
forward within a line & 54.4\% & 52.7\% & $-1.7$ & 0.005 \\
regression within a line & 15.1\% & 16.4\% & $+1.3$ & 0.018 \\
refixation & 12.8\% & 14.2\% & $+1.4$ & 0.061 \\
median amplitude, all & 2.74$^\circ$ & 2.69$^\circ$ & $-0.05^\circ$ & 0.21 \\
median amplitude, forward & 3.14$^\circ$ & 3.05$^\circ$ & $-0.09^\circ$ & 0.012 \\
\bottomrule
\end{tabular}
\end{center}

\noindent
The providers' component analysis did not remove all ocular activity. In the model with saccade
covariates, the kernel for signed horizontal displacement differed between F8 and F7 by
$+3.4$\,\si{\micro\volt} per standard deviation of displacement ($p=4\times10^{-15}$), the
signature of the corneo-retinal dipole. We modeled the saccade rather than rely on the
cleaning. The rescaling test under each model:

\begin{center}
\small
\begin{tabular}{lrrr}
\toprule
Model & Attenuation & 95\% CI & Difference field $p$ \\
\midrule
published design & 42.7\% & $[21.6, 73.1]$ & $2\times10^{-4}$ \\
+ amplitude spline and saccade type & 42.1\% & $[19.8, 68.3]$ & $2\times10^{-4}$ \\
+ signed horizontal and vertical displacement & 42.4\% & $[18.9, 67.9]$ & $2\times10^{-4}$ \\
same, state effect for forward saccades only & 39.7\% & $[11.5, 67.9]$ & 0.007 \\
\bottomrule
\end{tabular}
\end{center}

\noindent
Intervals in this table come from their own bootstrap draw and differ slightly from those in the
main text. The frequency-by-state interaction was similar ($+57\%$, $[-18, +149]$, $p=0.19$). Matching on amplitude tests the question without a model of the amplitude effect. Within reader,
we binned forward saccades by amplitude in 0.25$^\circ$ steps. We drew equal numbers of
mind-wandering and on-task fixations in each bin. This keeps 99\% of forward mind-wandering
fixations and makes the median amplitude equal (2.9$^\circ$ in both). We estimated the state effect on the matched sets.
A matched estimate depends on which on-task fixations are drawn, and we therefore repeated the
draw ten
times and compared the result with ten random subsets of the same size drawn without regard to
amplitude. Matched draws averaged 49.4\% (range 40.9 to 59.4\%), and random draws 45.9\% (35.8 to
52.8\%; difference $p=0.34$). Matching on saccade type and amplitude together over all fixations yielded 35.8\% against 39.4\% for random subsets (eight draws each; $p=0.38$). Matching removes nothing beyond the effect of subsampling. Single draws vary by about six points,
and one early draw of each kind was low (25.5\% and 19.9\%). For this reason, we report
distributions.

\paragraph{When the loss appears.} We refitted the deconvolution with time from the marked onset
entered as a set of indicator kernels (model A) and, separately, time to the span offset (model
B), with frequency, length, surprisal, log duration, incoming saccade amplitude and position in
the page as covariates. We projected the 150 to 290\,ms map of each bin on the leave-one-reader-out on-task field. A value
of 0 is no change, and $-0.5$ is a response at half its size. We dropped bins with fewer than 20
fixations in a reader for that reader.

\begin{center}
\small
\begin{tabular}{lrrrr}
\toprule
Bin & Proportional change & 95\% CI & $p$ & Readers \\
\midrule
$-20$ to $-10$\,s & $+0.11$ & $[-0.11, +0.34]$ & 0.353 & 43 \\
$-10$ to $-5$\,s & $-0.13$ & $[-0.40, +0.13]$ & 0.334 & 43 \\
$-5$ to 0\,s & $+0.19$ & $[-0.04, +0.46]$ & 0.139 & 44 \\
0 to 2\,s & $-0.17$ & $[-0.47, +0.11]$ & 0.273 & 42 \\
2 to 4\,s & $-0.29$ & $[-0.64, +0.08]$ & 0.112 & 39 \\
4 to 7\,s & $-0.66$ & $[-1.07, -0.26]$ & 0.004 & 40 \\
7 to 12\,s & $-0.34$ & $[-0.83, +0.15]$ & 0.202 & 37 \\
beyond 12\,s & $-0.84$ & $[-1.53, -0.17]$ & 0.024 & 25 \\
10\,s before onset (model B) & $+0.04$ & $[-0.11, +0.19]$ & 0.641 & 44 \\
more than 12\,s before offset & $-0.30$ & $[-0.88, +0.29]$ & 0.323 & 26 \\
12 to 6\,s before offset & $-0.37$ & $[-0.82, +0.02]$ & 0.090 & 39 \\
6 to 3\,s before offset & $-0.17$ & $[-0.44, +0.11]$ & 0.238 & 40 \\
last 3\,s before offset & $-0.47$ & $[-0.76, -0.18]$ & 0.004 & 43 \\
\midrule
\multicolumn{5}{l}{\emph{After the offset (model C)}} \\
reading after the span, $>$5\,s before the report & $+0.36$ & $[-0.25, +0.98]$ & 0.267 & 26 \\
last 3\,s before the report & $-1.02$ & $[-1.52, -0.54]$ & 0.0003 & 40 \\
control: last 3\,s before a page advance & $-0.02$ & $[-0.31, +0.26]$ & 0.870 & 44 \\
control: 3 to 6\,s before a page advance & $+0.07$ & $[-0.19, +0.32]$ & 0.581 & 44 \\
\bottomrule
\end{tabular}
\end{center}

\noindent
No reliable change appears before onset. The loss is reliable from 4 to 7\,s into a span and is
still present in the last three seconds before the offset. Model C examines what happens after the
offset. In the release, the labeled span ends at the offset, and first-pass reading ends at the
report. The interval between them is usually short (median 2.1\,s; 17.5\% of lapse pages have more
than 5\,s of reading after the offset). A single ``after the offset'' bin would contain
mostly the approach to the report, not reading. When we separate the two, reading after the span
and well before the report is at baseline. The last three seconds before the report show a loss
larger than the loss in the span.

A control shows which of these effects is related to the end of first-pass reading. On pages with
no lapse, the last three seconds before a page advance show no change (11{,}472 fixations from all
44 readers), as does the window before them. The recovery bin uses 26 readers,
and its interval allows a moderate residual loss. It bounds the persistence of the
effect but does not exclude it. The bins are small, the intervals are wide, and the marked onset
is the reader's own judgment. These estimates establish sequence, not a precise latency.

For these reasons, the main text treats this neural effect as a change of state. A proportional change in
amplitude applies equally to every word. It cannot show which particular word was
processed less, as a change in the frequency or surprisal kernel would. The neural
signature of mind-wandering is real, and it is uninformative about content by construction.

These are sensor-level results, and we do not interpret the scalp map as source localization.

\paragraph{Procedure.} We asked whether mind-wandering changes the pattern or the size of the
fixation-related response. For each reader, we took the deconvolved intercept kernel and the
mind-wandering indicator kernel of the model in Methods. This model has frequency, length,
surprisal and fixation duration as covariates (the saccade-controlled refits are above). Next, we tested the
difference field with a sign-flip test on global field power. We correlated its spatial
pattern with the on-task fixation field over the same interval and bootstrapped over readers.

The proportional model estimates a single scalar $\alpha$ such that the difference field equals
$-\alpha$ times the on-task field. We fit it leave-one-reader-out, so that no reader's own data enters their own estimate. We report
how much of the variance in the group difference map the rescaled on-task map accounts for. After
we subtract it, we test the residual field with the same sign-flip procedure. Time-resolved differences use a familywise cluster correction across
the epoch. We do not retain polarity when comparing topographies, because a polarity-preserving
comparison confounds a change in sign with a change in configuration.

\subsection*{S8. Distribution of additional viewing time}

We tested whether additional viewing during a lapse was concentrated on words with greater measured difficulty.
For each reader, we regressed a viewing measure on word difficulty, state and their interaction, and tested the interaction across readers. There were three viewing measures and three
difficulty measures (surprisal, rarity and length). Eight of the nine tests were not significant before correction; none was significant after Holm correction. The gradient on refixations changes by $+5.0\%$ of its on-task
value during mind-wandering (95\% CI $[-4.5, +14.4]$), and the gradient on fixation duration
changes by $-9.1\%$ $[-45.3, +25.6]$. These intervals constrain some large changes, but leave smaller changes unresolved.

The ninth test, regressions against rarity, is nominal ($p=0.024$, $p_{\text{Holm}}=0.22$). The dependence on rarity was estimated to weaken by 56\%, but this comparison did not survive correction. Nor do quartile splits of surprisal show a trend ($p=0.25$, $0.33$ and $0.54$). We also tested sensitivity to extended context. We compared surprisal computed over the whole
preceding passage with surprisal computed within the sentence alone. The interactions with the extended-context term were not significant for reading time ($p=0.37$) or the N400 ($p=0.28$). These tests do not establish equivalence or locate the comprehension failure.

We also tested a model-based prediction. We took the residual after the
geometry model and the on-task text model of S5. Next, we asked whether any word property predicts
the additional dwell during mind-wandering, with cross-validation by reader. The held-out $R^2$ is
$-0.005$ on mind-wandering transitions ($n=28{,}450$) and $-0.003$ on on-task transitions
($n=364{,}316$). This model did not predict the additional viewing from the tested text features.

The three viewing measures do, however, change differently over time. Fixation duration and refixation
change in a single step. They are fully present in the first two seconds of a span and do not
change after that (trends $p=0.63$ and $p=0.49$; against matched on-task stretches,
$p_{\text{perm}} =0.52$ and $0.43$). Regressions are different. The regression rate increases
steadily through a span, from $+0.016$ above the reader's own baseline in the first two seconds to
$+0.080$ after ten seconds (linear trend $+0.022$, 95\% CI $[0.011, 0.034]$, $p=0.0012$, 26 of 37
readers). This increase is specific to reported spans. In 300 matched on-task stretches with the
same length distribution, no trend was this large ($p_{\text{perm}}<0.005$). The trend remains
when we remove each reader's position through the run ($+0.022$, $p=0.0007$), so adjustment for measured position through the run did not remove the trend.

Regression behavior, unlike fixation duration or refixation, accumulates across a span. The increase reflects backward eye movements; the tested word-property interactions did not identify difficulty-specific targeting.

\paragraph{Procedure.} Here, a span is a maximal run of consecutive fixations with a
mind-wandering label. For each fixation, we recorded the time since the first fixation of its span
and binned it into 0--2, 2--5, 5--10 and beyond 10 seconds. We expressed each reader's values
relative to their own on-task mean and fitted a linear trend across bins for each reader. The null
came from 300 sets of matched on-task stretches. We drew them at random positions, sampled their
lengths from the observed distribution of span lengths, and analyzed them identically. We
controlled for time on task by regressing each reader's regression indicator on their position
through the run before binning.

For the targeting analysis, we fitted each viewing measure for each reader on standardized
difficulty, state and their interaction by least squares. We tested the nine interaction
coefficients across readers with Holm correction. Difficulty was surprisal, negative Zipf
frequency and length in turn. The quartile version splits words into quartiles of surprisal
within reader and fits a linear trend across the four differences. To test extended-context integration, we decomposed GPT-2 surprisal into a within-sentence
component and the additional reduction from the preceding passage and tested the interaction
of that second component with mind-wandering.

\subsection*{S9. Ongoing state: alpha, pupil, beta and inter-subject alignment}

\paragraph{Band power across readers.} Section~\ref{sec:alpha} reports one spectral effect
shared across readers. The per-reader tests are liberal, since segments within a span are
autocorrelated. Readers needed at least five complete 2\,s segments in each state, which one
reader lacked in both variants and two more lacked in the clean one. Per-reader significant channels were common in every band. At least one FDR-significant channel
occurred in 20 of 43 readers for theta, 25 for alpha, 28 for beta and 32 for gamma. Their
directions did not agree. The mean correlation between two readers' effect maps was
0.007, 0.033, 0.016 and 0.023 for the four bands. Only alpha produced channels that are significant under a sign-flip max-$t$ test over all 64
channels. In the naive contrast there were 18 such channels, led by P7 ($+14.0\%$, 37 of 43
readers), P5, TP7, P9 and P3, and extending over left centro-parietal and fronto-central sites. In
the clean contrast there were four: P7, P9, P5 and TP7. Theta fell over centro-parietal sites by 5 to 7\%, and beta and gamma over posterior sites by 3 to 5\%. Yet no channel was significant in any of the three bands in the naive
contrast, and only one channel each for theta and beta in the clean contrast.

\paragraph{Alpha after removing fixation-evoked activity.} For each reader, we fitted a deconvolution model to every
fixation, first-pass and rereading. The model had kernels for the intercept, first-pass status and
mind-wandering state over $-100$ to $+600$\,ms, with the per-column penalty of the main analysis.
We subtracted its prediction from the recording. The state kernel removes the change in the
fixation-related response during mind-wandering, together with the change in fixation timing. The
evoked prediction contains 3.5\% of the variance of the recording. We repeated the segment analysis on the raw recording, on the residual and on the prediction alone. We did this in the
naive and clean variants, for the 8 to 13\,Hz band and for a narrow 9.5 to 10.5\,Hz band.

We set the criterion before running the test. The cluster effect had to keep at least 70\% of its
size in the residual, and P7 had to stay significant under the max-$t$ correction. The residual
kept 90.4\%, 100.9\%, 91.5\% and 97.9\% of the cluster effect in the four variants. P7 stayed
significant in each ($p\le0.006$). The residual and raw topographies correlated at 0.96 to 0.98,
and occipital alpha remained null ($p=0.51$ to $0.90$).

\paragraph{Alpha and the rescaling.} At the five channels of the unexcluded alpha cluster, the on-task fixation field is
$-0.224$\,\si{\micro\volt}, and its change during mind-wandering is $+0.075$\,\si{\micro\volt}.
This is a loss of 33\%, no larger than the 43\% whole-scalp rescaling. Over the 64 channels, the
alpha change map does not follow the magnitude of the fixation field ($r=-0.11$, $p=0.37$). Across
readers, the alpha increase at the cluster does not predict the rescaling ($\rho=-0.15$, $p=0.33$,
43 readers; $\rho=-0.29$, $p=0.067$, 41 readers in the clean variant). The negative estimate is consistent with an inhibitory interpretation, but is too imprecise to establish a relationship.

\paragraph{Alpha on the answer span.} We fitted the model of Section~\ref{sec:comp} on 1804 trials
from 44 readers. Alpha on the answer span and elsewhere on the page entered alongside page
coverage. Left posterior-lateral alpha on the span was associated with $-0.4$ percentage points of accuracy per SD (95\% CI $[-2.9, +2.1]$,
$p=0.75$). With the mind-wandering terms added, $p=0.83$, and the report on the span held at 25.8 points for a lapse over the whole span ($p=3.1\times10^{-8}$). On the 1498 trials
with no reported lapse on the span, $p=0.52$. The right-hemisphere homologue showed $+2.5$ percentage points of accuracy
per SD ($p=0.041$). This direction is opposite to an inhibitory account, and we did not predict
it. It is not distinguishable from its own random-set null (two-sided $p=0.063$). The occipital control was null ($p=0.47$).

\begin{figure}[!tbp]
\centering
\includegraphics[width=0.78\textwidth]{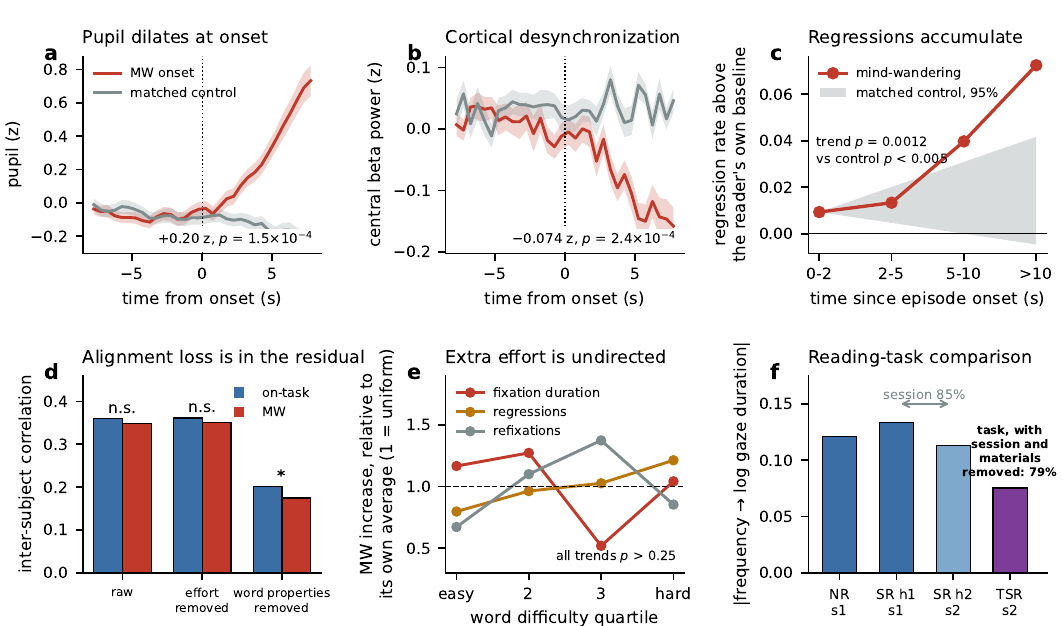}
\caption{\textbf{What changes during mind-wandering, and the positive control.}
(a, b) Pupil diameter and central beta power time-locked to the onset of a reported span,
against position-matched control onsets. Shading is the standard error across readers.
(c) Regression rate through a span, relative to each reader's own on-task baseline. The band
is the 95\% range of trends from 300 matched on-task stretches of the same length distribution.
(d) Inter-subject alignment against a leave-one-reader-out template, before and after removing
the additive shift and the word-property component.
(e) The mind-wandering increase in each viewing measure across quartiles of word difficulty,
scaled to its own average. A flat line means the increase is uniform.
Panels a to e are the primary sample, 44 readers. (f) Frequency coupling in the independent
sample of 12 readers by condition, with session (s1, s2). The arrow is the session effect between the two halves of the sentiment task. The task effect
removes both session and materials (S10).}
\label{fig:changes}
\end{figure}

\paragraph{Pupil and beta at span onset.} Time-locked to the onset of a reported span, pupil diameter increased ($+0.20$\,z,
$p=1.5\times10^{-4}$) and central beta power decreased ($-0.074$\,z, $p=2.4\times10^{-4}$).
The matched on-task controls did not show either change
(Fig.~\ref{fig:changes}a--c). The literature associates both changes with mind-wandering outside
reading \citep{smallwood2011pupil,kam2011,braboszcz2011}. These trajectories extend to 8\,s after
onset, and many reported spans are shorter than that. For those spans, the window after onset
includes the approach to the report. We cannot attribute these two changes to the lapse
itself rather than to the reader noticing it.

Readers also became less similar to one another, in one component of gaze behavior only. We
built a leave-one-reader-out template from other readers' on-task gaze durations for the same
word. Against this template, overall alignment changed very little (retention 97.1\%, $p=0.30$).
When we removed the additive and scale shift by z-scoring within state, the result was the same
(97.0\%, $p=0.29$). Word properties account for much of the shared variation; the corresponding alignment contrast was not significant.

The effect appears only after that component is removed. In the residual after regressing out
frequency, length and surprisal, alignment dropped from 0.201 to 0.175 (retention 86.7\%,
$\Delta=-0.027$, 95\% CI $[-0.051, -0.002]$, $p=0.037$). Shuffling the mind-wandering labels
within reader at matched count never reproduced it ($p_{\text{perm}}=0.007$). Against a raw group
template instead of a leave-one-reader-out template, the same contrast gives $\Delta=-0.028$
($p=0.025$). The two estimates agree in direction, and the difference in size comes from how the
template is built (Procedure, below).

The reliable alignment difference appeared after removing the modeled word-property component. Readers retain the measured word-property associations, while the remaining portion of
their behavior diverges.

\paragraph{Procedure.} We computed peri-onset trajectories over $\pm 8$\,s around the onset and
offset of each reported span, z-scored within run, for pupil diameter and for band-limited power.
We drew control onsets from on-task reading, matched on position within the run.
Inter-subject correlation used a leave-one-reader-out group template of the response to each
word. For a given reader, that template is the mean over the other readers' \emph{on-task} readings of
that word, with at least five contributors. It represents a normative response, not a
mixture of states. We correlated the target reader's words with the
template separately for each state, subsampling on-task words to match the mind-wandering
count. The response took three forms: raw log gaze duration, the same z-scored within
reader and state, and the residual after regressing out frequency, length and surprisal. We
tested specificity by permuting the mind-wandering labels within reader at matched count.

 \paragraph*{Part V. The control sample.}

\subsection*{S10. The control sample: session, materials and skipping}

\begin{figure}[!tbp]
\centering
\includegraphics[width=\textwidth]{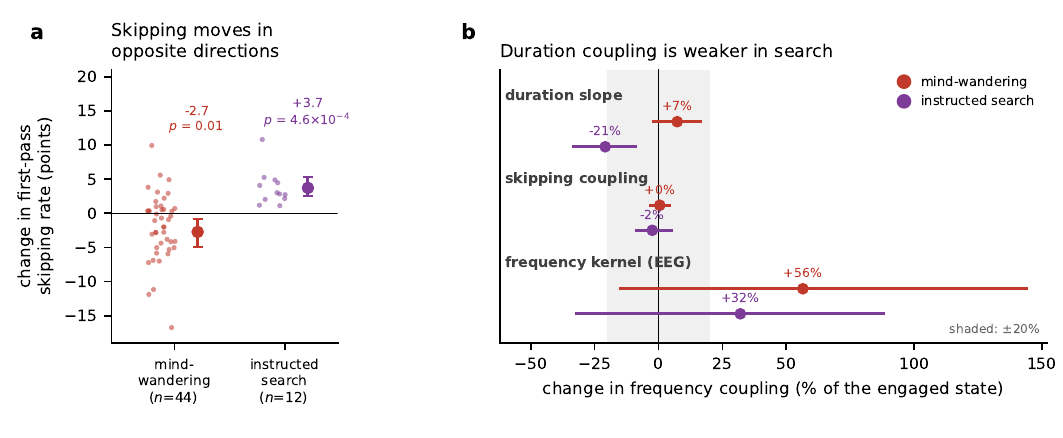}
\caption{\textbf{Semantic-relation search and mind-wandering change reading in different ways.}
(a) Change in the first-pass skipping rate, one point per reader, with the mean and its 95\%
bootstrap interval. Mind-wandering: primary sample, 44 readers, steps within a text line.
Instructed search: control sample, 12 readers, the task effect with session and materials
removed.
(b) Change in frequency coupling as a percentage of the on-task or normal-reading value, with 95\%
intervals. Duration slope: for mind-wandering, the within-token interaction with reader and token
fixed effects; for instructed search, the task effect with session and materials removed. Skipping
coupling: Somers' $D$. Frequency kernel: the deconvolved occipitotemporal response, shown for
comparison. Both neural intervals are wide. Shading marks $\pm$20\%.}
\label{fig:states}
\end{figure}

\paragraph{Task comparison summarized in the main text.}
The independent sample tested whether the measures could detect a known change in reading goal
(Fig.~\ref{fig:states}). Readers read encyclopedic sentences either for comprehension or while
searching for a named semantic relation. Since task order was fixed, the two conditions also differ
in recording session. A third task, sentiment reading of film reviews, ran across both sessions.
We therefore defined the task effect as the change from normal reading to search minus the change
between the two halves of the sentiment task. This difference removes both session and materials
(SI~S10).

The instructed search made reading shallower. Total reading time fell by 20\%, readers
fixated 31\% fewer words, and forward steps lengthened from 1.71 to 1.96 words. First-pass skipping rose by 3.7 points (95\% CI $[2.5, 5.3]$, $p=4.6\times10^{-4}$, 12 of 12 readers). This is
the opposite of the change during mind-wandering (Fig.~\ref{fig:states}a). The frequency slope on gaze duration dropped to 79\% of its value in normal reading (reader-bootstrap 95\% CI $[67, 91]$; $t$-test on the per-reader slope change, $p=0.009$), and the length slope to 82\% ($p=0.012$). The surprisal slope, at 88\%, was not reliably reduced
($p=0.16$).

Skipping coupling did not change. The task left it at 98\% of its normal-reading value for
frequency, 101\% for length and 94\% for surprisal (all $p\ge0.085$), and the first-pass
definition gave the same result (SI~S10). The instructed readers skipped more words, but their
choice of which words to skip was as strongly coupled to word properties as before. Skipping
coupling, then, does not separate normal from shallow reading. Its preservation during
mind-wandering is informative only together with the duration result. The deconvolved frequency
kernel changed by $+32.2\%$ ($[-30.6, +88.4]\%$, $p=0.33$, $n=12$). With twelve readers, this is a
weak constraint.

The control sample differs from the primary one in laboratory, sample size, materials and hardware
as well as in goal, so we use it for one thing only: to show that the duration measure can detect
a change in coupling (Fig.~\ref{fig:states}b).

\paragraph{Session, task and materials.} Task order in that dataset was fixed for every reader, and the two encyclopedic tasks were in
different sessions \citep{hollenstein2018}. The
sentiment task ran across both sessions, and we split it at its sentence-index midpoint. Each reader then contributes four cells: normal reading (session 1, encyclopedic),
sentiment first half (session 1, film reviews), sentiment second half (session 2, film reviews)
and relation search (session 2, encyclopedic). The task effect is (search $-$ normal reading)
$-$ (second half $-$ first half), which removes session and materials together. The split assumes that sentence index follows presentation order. The dataset documentation
supports this assumption, although we could not verify it directly.

There is a session effect: frequency coupling on gaze duration fell to 84.9\% between the two
halves of the sentiment task ($p=0.020$). There is also a materials effect: film-review sentences
had stronger coupling than encyclopedic sentences (110.5\% for frequency, 134.2\% for surprisal).
A comparison of search with the second sentiment half gives 66.4\% for frequency and 56.0\% for surprisal, but changes both task and materials. With session and materials removed, the task
reduced the frequency slope to 79.1\% ($[66.7, 91.1]$, $p=0.009$, 9 of 12 readers), the length
slope to 82.0\% ($[70.4, 92.9]$, $p=0.012$) and the surprisal slope to 87.8\% ($[72.0, 103.6]$,
$p=0.16$).

\paragraph{Skipping.} We scored skipping in these single-line sentences with the rules of S4,
ordering words by their first fixation. With a step limit of four words, the task increased first-pass skipping from 0.393 to 0.450
against normal reading. With session and materials removed, the increase was 3.7 points ($[2.5,
5.3]$, $p=4.6\times10^{-4}$, 12 of 12 readers). Forward steps lengthened from 1.71 to 1.96 words. The rise held at every step limit from one to ten words. Skipping coupling did not change.
With session and materials removed, retention under the never-fixated rule was 97.6\% for
frequency ($[91.5, 104.9]$, $p=0.52$), 100.6\% for length ($[94.8, 106.9]$, $p=0.85$) and 93.7\%
for surprisal ($[88.4, 100.0]$, $p=0.085$). Under the first-pass rule, it was 102.5\%, 105.2\% and
93.3\% (all $p\ge0.11$). The same 48 sentences read under both tasks
gave 90.9\% ($p=0.061$), 103.7\% ($p=0.44$) and 89.9\% ($p=0.035$), with session not removed. On
task, skipping rates in the two datasets were similar under the same rule (ROAMM within a line
0.372, ZuCo normal reading 0.393; $p=0.47$).

\paragraph{The neural comparison is weak.} The occipital frequency kernel changed by $+32.2\%$
($[-30.6, +88.4]\%$, $p=0.33$, $n=12$) under the instructed goal, against a behavioral
attenuation of 12 to 21\%. With twelve readers, this is a weak constraint. The main text treats it as agreeing in direction
with the primary-sample frequency result, not as an independent finding.

\paragraph{A dissociation we looked for and did not find.}

We tested for a selective dissociation between lexical and semantic attenuation under the shallow
goal. The data did not support one, and we report none. Collinearity between the two predictors produced the apparent effect. The bivariate estimates show
about equal attenuation.

\paragraph{A scale artifact in how coupling is reported.}

A scale artifact is common in this literature. When coupling is expressed as a slope on log
duration, a purely additive change in duration rescales the slope by the change in the mean.
Mind-wandering lengthens gaze durations by 16\% pooled over words. An unchanged coupling would appear as 86.2\% retention on the log scale. Instructed search makes reading 11\%
faster, so an unchanged coupling would appear as 112.0\%. A log-scale slope is biased
toward attenuation under an additive slowing. A preserved log slope during mind-wandering is
therefore a conservative result. On the raw millisecond scale, mind-wandering retention is 133.8\% and
instructed-search retention is 53.4\%.

\paragraph*{Part VI. Comparison with related reports, and robustness of the comprehension
results.}

\subsection*{S11. Decomposing the reported loss of brain--language alignment}

 \citet{sun2026alignment} analyze the same recordings with an encoding model and report that
alignment between language-model embeddings and fixation-aligned electroencephalography is
reduced during mind-wandering, strongly for spectral power and weakly for fixation-related
potentials. Section~\ref{sec:bridge} asks whether the proportional rescaling of S7 already explains the
fixation-related part of that reduction. The full decomposition follows.

\paragraph{The identity.} Encoding accuracy is a Pearson correlation between predicted and
observed responses. Writing $\beta$ for the least-squares slope of observed on predicted,
$r = \beta\,\sigma_{\text{pred}}/\sigma_{\text{obs}}$ exactly, so
\[
\frac{r_{\text{MW}}}{r_{\text{on}}}
= \underbrace{\frac{\beta_{\text{MW}}}{\beta_{\text{on}}}}_{\text{response gain}}
\times \underbrace{\frac{\sigma_{\text{pred,MW}}}{\sigma_{\text{pred,on}}}}_{\text{stimulus}}
\times \underbrace{\frac{\sigma_{\text{obs,on}}}{\sigma_{\text{obs,MW}}}}_{\text{noise}} .
\]
The first term describes the fitted response amplitude. The other terms describe predicted and observed dispersion and can also depend on neural signal and noise. Uniformly multiplying the entire observed signal leaves Pearson correlation unchanged; a response-component rescaling relative to other activity can change alignment. This identity is descriptive and does not by itself identify a mechanism or establish unchanged representations.

\paragraph{Procedure.} We reduced each fixation-locked epoch to its mean amplitude over 200 to
300\,\si{\milli\second} at each of 64 sensors. Fixation-related alignment is highest in this
window. The baseline was $-100$ to $0$\,\si{\milli\second}. Within each reader, we paired all
first-pass mind-wandering fixations with an equal number of on-task fixations drawn at random.
This yielded 29,201 fixations in each state from 43 readers. We divided each reader's amplitudes by
that reader's on-task standard deviation, so that the between-state amplitude ratio under test is
not changed. The predictors are GPT-2 hidden states for the fixated word, reduced by principal
components fitted within the training fold.

We pooled a single ridge encoder across readers, as per-reader fits are at the noise floor
and resolve nothing (cf.~S6). We chose the regularizer on an inner split. We cross-validated
five-fold with folds defined by page, ensuring that no page contributes to both training and evaluation. We
selected sensors as the top quartile by on-task alignment, a criterion blind to the state
contrast.
Intervals come from 2000 bootstrap resamples over readers.

\paragraph{Result.} The main numbers come from the specification with the highest on-task
alignment, the sixth GPT-2 layer at 50 components. That rule uses on-task performance only and is
blind to the state contrast. On-task alignment was $r=0.0277$, and mind-wandering alignment was
$r=0.0182$, a ratio of $0.657$ (95\% CI $[0.379, 1.020]$). The difference of $0.0095$ has an
interval of $[-0.0005, 0.0196]$. The interval includes no difference. The decomposition itself is
better resolved. The response-gain term was $0.662$ (95\% CI $[0.369, 1.046]$), the stimulus
term $1.014$ and the noise term $0.961$ (95\% CI $[0.894, 1.017]$).

The identity is exact at each sensor. Each term is averaged over the selected sensors separately,
so their product, $0.645$, differs slightly from the ratio of the averaged correlations. The
recovered response-gain ratio is statistically indistinguishable from the $0.573$ remaining
response from the leave-one-reader-out rescaling fit of S7 (bootstrap $p=0.60$). That fit uses a
different estimator on different features. The single-penalty remaining-response estimate is $0.742$ and is also compatible with the encoding ratio ($p=0.66$). The wide interval cannot distinguish these estimates.

\paragraph{Robustness.} We ran all seven cached GPT-2 layers at 16 and 50 components, fourteen
specifications in total. Alignment was lower during mind-wandering in every specification, with ratios from $0.352$ to
$0.689$. The response-gain term was within $0.02$ of the ratio in every specification. The stimulus term ranged from $0.968$ to $1.034$, and the noise term from $0.958$ and
$0.973$. The conclusion does not depend on the layer or the dimensionality.

\paragraph{Limits of the weight comparison.} We also compared the encoding weights fitted separately in
each state. That comparison is uninformative. The within-state split-half reliability of the weight maps is itself low. It ranges from $0.68$
down to zero across specifications, and it is slightly below zero for mind-wandering in two of
them. A between-state similarity thus has no stable reference value.

The evidence for proportional scaling of the measured field comes from S7. S7 tests proportional scaling with 5000 permutations and finds no reliable residual; this does not establish exact equality of the underlying source configurations. We make no claim about the spectral half of their result. A model fitted to alpha and low-beta power uses bands that index the
mind-wandering state directly: our own data show a decrease in central beta at span onset (S9). We
have not separated that effect from a change in what those bands encode about the text.

 \subsection*{S12. Comprehension: answer spans, error composition and effect size}

\paragraph{Validating the answer spans.} The primary annotation marks a median of 21\% of the
page. The second annotation agrees closely. It has identical sentence sets for 32 of the 50 items and
the same item type for 45. The mean Jaccard overlap is 0.825, sentence-level Cohen's $\kappa$ is
0.83, and the two annotations do not overlap at all on one item.

We used Qwen2.5-1.5B-Instruct to
check that each annotation contains the answer. We scored the four option letters under eight
random orderings of the options and averaged the probability on the keyed option. With only the
primary spans, the model assigned 0.892 to the keyed option and chose it on 89.8\% of items. The corresponding
values were 0.859 with the spans of the second annotation, 0.848 (85.0\%) with the whole page, and
0.545 (55.8\%) with no passage. When we recomputed the gaze null of Section~\ref{sec:comp}
under the second annotation, the true answer span was at the 33rd percentile of the random-set
distribution ($p=0.67$). This is the same as under the primary annotation. Both annotations are
automatic, and we have not measured agreement between human annotators.

\paragraph{Primary and alternative comprehension estimates.}
The primary model regresses accuracy on the fraction of answer-span fixations labeled as mind-wandering, with reader and item fixed effects and covariates for the lapse fraction elsewhere and page coverage. The coefficient is $-0.256798$ (SE $0.046891$; reader-clustered normal-Wald 95\% CI $[-0.348703,-0.164893]$, $p=4.3388\times10^{-8}$). Thus the full-span versus no-span contrast is a 25.7-point loss. Re-estimating this model from the stored trial table reproduces the saved coefficient to numerical precision.

The following alternatives answer different questions. Binary coding asks whether any lapse reached the answer span: its estimated loss is 18.3 points ($[11.6,25.0]$, $p=8.4\times10^{-8}$). A fraction-based model without page coverage gives 30.8 points for full-span versus no-span coverage ($[22.6,39.0]$, $p=1.7\times10^{-13}$). Adding answer-span gaze coverage, fixation count and dwell time to the adjusted model gives 22.4 points ($[12.8,32.0]$, $p=4.6\times10^{-6}$). None of these is an additional primary result. Raw accuracy is 0.680 with no lapse, 0.640 with a lapse elsewhere only, and 0.405 with a lapse on the answer span.

\paragraph{Random sets and the fixed non-answer control.}
Random controls sample sentences in random order from the same page, stopping at the answer span's word count and truncating the final sentence as needed. These sets may be noncontiguous and may include answer-bearing text. The fixed control was selected once, using seed 60, from sentences outside the answer span and outside the separately marked contradicting sentence where applicable. Its word count matches the answer span for 48 of 50 items. The remaining two items contain too little eligible text: Prisoners Dilemma\_p1 has 57 control words versus 119 answer words, and Prisoners Dilemma\_p4 has 89 versus 137. Its 0.9-point estimated loss is therefore a check using non-answer content, not an estimate of the random-set median (10.8 points).

The overlap gradient summarizes 1000 further sentence-sampled draws, using seed 60888. The fitted slope corresponds to 3.4 accuracy points per additional 10\% overlap. The draws reuse the same readers and items; a regression $p$ value across these draws is not an independent participant-level test. The random-set comparison in the main text is the inferential test of content specificity.

\paragraph{The mis-keyed item.} One item (Prisoner's Dilemma, page 8) keys an option the page
contradicts, and the page states another option verbatim. Its 39 trials are in the models of
Section~\ref{sec:comp}. Without them ($n=1765$), a lapse on the answer span lowered accuracy by 19.2 points
($p=5\times10^{-8}$), and a lapse over the whole span lowered it by 31.9 points
($p=3\times10^{-13}$). In the model that adds page coverage, the whole-span loss was 26.6 points,
against a median of 11.3 for random regions of the same size. Only 2 of 1000 random regions
matched or exceeded it ($p=0.003$). The attributable share of errors was 5.4\%.

\paragraph{Uncertainty responses.} The rate of ``not sure'' answers was estimated to rise by 8.6 points, but the association was not significant ($p=0.099$). This result does not establish whether readers recognized their comprehension failures.

\paragraph{Effect size in proportion.} The attributable share of errors reported in
Section~\ref{sec:comp}, 5.1\%, has a 95\% CI of $[2.7, 7.7]$. The raw contrast between 0.680 and 0.405 overstates the population
effect considerably.

\paragraph{Answerability without the passage.} Section~\ref{sec:comp} states the no-passage answerability of the item bank. This model-based check indicates that some items can be answered without the text; it is not a human performance ceiling. Item identity is held constant in the content-specificity comparison.

\subsection*{S13. Rebuilding the global gaze model of comprehension}

 \citet{southwell2020} predict comprehension, assessed after reading a long connected text, from
seven global page-level gaze features. They use participant-level cross-validation and report
observed-versus-predicted correlations of $0.384$, $0.362$ and $0.372$ in samples of 104, 130 and
147 readers. Section~\ref{sec:comp} uses a rebuild of that model on our readers to show the level
at which our gaze null applies. We describe that rebuild below.

\paragraph{Features and exclusions.} We computed their seven features per reader and page: mean
fixation duration, number of fixations, regression fixation proportion, mean saccade length,
horizontal saccade proportion (saccades within $30^\circ$ of horizontal), fixation dispersion
(root mean square distance of each fixation from the page's mean fixation) and reading time. We identified regressions from screen coordinates, not from word indices. A regression is a
fixation that lands left of the previous one on the same line, or on an earlier line. This form of
the feature is robust to tracking errors, which is the reason that they chose global features. We applied their exclusions. We dropped pages with less than one second of reading or at most two
fixations. We truncated feature values beyond 2.5 median absolute deviations to the extreme value
within that bound. This leaves 2,187 pages from 44 readers, with an observed accuracy of
0.620.

\paragraph{Model and metric.} We fitted logistic regression on page-level accuracy, with
participant-level four-fold cross-validation. No reader is in both training and evaluation. The metric is theirs: the correlation, across held-out readers, between each
reader's observed mean accuracy and their mean predicted probability. We repeat the whole
procedure 100 times with different fold assignments and report the median, as they do.

\paragraph{Result.} The median held-out correlation is $0.338$ (range across repetitions
$0.148$ to $0.468$, median $p=0.025$), against $0.384$, $0.362$ and $0.372$ in their three
samples. With a shuffled outcome, it decreases to $0.009$. Reading time alone gives $0.108$ and is not
reliable ($p=0.46$). This agrees with their finding that the model is not a reading-time
model. Their result replicates on these readers.

\paragraph{Their head-to-head test.} \citet{southwell2020} then regress observed comprehension on the model's prediction and on
self-caught mind-wandering together, between readers. They find that the gaze prediction is
reliable ($\beta=0.40$, 95\% CI $[0.23, 0.57]$) and that mind-wandering is not ($\beta=0.13$,
$[-0.04, 0.30]$, $p=0.13$). On our 44 readers, the same test gives
$\beta=0.24$ ($[-0.15, 0.63]$, $p=0.22$) for the gaze prediction and $\beta=-0.16$ ($[-0.54,
0.23]$, $p=0.42$) for the reader's mind-wandering rate, with $R^2=0.130$. Neither term is
reliable. We have 44 readers against their 130, and this test uses only the between-reader variance in
accuracy. We treat it as uninformative, not as a replication or a contradiction of their
ordering. The replication that we claim is of the model, above. We report
it to make the comparison with their study explicit.

\paragraph{The level at which the records separate.} Within reader, after we remove each reader's own mean, both records contain page-level
information. The gaze model's prediction correlates with page accuracy at $r=0.130$
($p=1.1\times10^{-9}$), and a reported lapse on the page correlates at $r=-0.101$
($p=2.2\times10^{-6}$). The separation appears one level lower, within the page. There, the effect
of a reported lapse on the answer span exceeds 992 of 1000 size-matched control regions, whereas gaze on the same span is at the 33rd percentile of that null (Section~\ref{sec:comp}).
Table~\ref{tab:levels} shows the three levels. We interpret this as the resolution of an apparent
conflict, not as a disagreement. They claim that eye movements predict comprehension, and we
reproduce this. We claim that eye movements do not localize the loss, and their design did not
test this.

\begin{table}[h]
\centering
\caption{\textbf{Gaze and lapse reports at three levels of analysis.} Each row is a different level of
analysis. Gaze contains the coarse, rate-like information and loses it as the question becomes
content-specific. The reader's own report does the opposite. Row 1 rebuilds
\citet{southwell2020} on these data: the gaze entry is that model's cross-validated correlation across held-out readers. The lapse
entry is the mind-wandering coefficient from their head-to-head regression, which is underpowered
with 44 readers. Row 3 is
Fig.~\ref{fig:comp}c.}
\label{tab:levels}
\small
\begin{tabular}{llll}
\toprule
Question & Level & Gaze record & Reported lapse \\
\midrule
Who comprehends more? & between readers & $r=0.338$ & $\beta=-0.16$ ($p=0.42$) \\
Which page went badly? & within reader & $r=0.130$ & $r=-0.101$ \\
Which content was lost? & within page & 33rd pct.\ of null & beats 992 of 1000 \\
\bottomrule
\end{tabular}
\end{table}

\subsection*{S14. The return window}

A corrective return is scored inside a window, and the window length is an analytic choice. The evidence below explains the five-fixation window used in Section~\ref{sec:skip}. It shows that
the mind-wandering contrast increases as the window becomes wider, and that this increase is an artifact of unequal opportunity, not evidence of successful repair. All values below come from
80{,}812 single-skip events in 44 readers. We scored them on the fully word-mapped fixation
sequence and confined them to a single page.

\paragraph{When returns occur.} Almost all returns to a stepped-over word are immediate. Only
23.0\% of stepped-over words receive a return at all. Of those returns, 55.6\% occur on the next
fixation, and the median occurs one fixation and 0.30\,s after the step. Cumulative capture is
55.6\% at one fixation, 70.3\% at three, 75.2\% at five, 82.8\% at ten, 94.1\% at twenty and
97.9\% at forty. The distribution has a sharp mode at the first fixation and a long thin tail. A
short window is thus not an arbitrary cut through a smooth distribution.

\paragraph{The contrast depends on the window.} When the window becomes wider, the mind-wandering
effect changes from null to significant.

\begin{center}
\small
\begin{tabular}{lrrrr}
\toprule
Window & On task & Mind-wandering & Change & $p$ \\
\midrule
1 fixation   & 0.1240 & 0.1314 & $+5.9\%$  & 0.37 \\
3 fixations  & 0.1564 & 0.1663 & $+6.3\%$  & 0.28 \\
5 fixations  & 0.1672 & 0.1824 & $+9.1\%$  & 0.16 \\
10 fixations & 0.1849 & 0.1991 & $+7.7\%$  & 0.19 \\
20 fixations & 0.2108 & 0.2379 & $+12.8\%$ & 0.030 \\
40 fixations & 0.2192 & 0.2525 & $+15.2\%$ & 0.010 \\
rest of page & 0.2240 & 0.2631 & $+17.5\%$ & 0.0038 \\
\bottomrule
\end{tabular}
\end{center}

\noindent
The change column is the reader-level mean difference over the on-task mean. The return-rate change summarized in Section~\ref{sec:skip} and the caption of Fig.~\ref{fig:measurement} ($+12.6\%$, $p=0.13$) differs from the five-fixation row for two reasons. It uses the mean of
each reader's ratio, as the other within-reader changes in that section do. Nor does it confine the lookahead to one page. Neither estimate is
reliable.

\paragraph{Why a window counted in fixations is not matched across states.} Fixations are not exchangeable between states, and the mismatch goes in two directions at the same
time. We count back in the fixations that have no word identity (off-text gaze and re-reading of
words already visited). The nominal five fixations then span 7.15 fixations and 1.81\,s of
recording on task, against 12.07 fixations and 3.21\,s during mind-wandering (difference $+1.92$\,s
$[1.05, 3.03]$, $p=5.3\times10^{-4}$, 34 of 44 readers). In effect, mind-wandering gets about twice
the time to produce a corrective return. In the other direction, single skips during
mind-wandering occur later on the page. The median event has 27 fixations and 7.1\,s of page
remaining, against 103 fixations and 26.0\,s on task. A wider window changes the second imbalance
into the first.

\paragraph{The matched test.} A window defined in seconds, with only the events that have at least
that much page remaining, controls both imbalances. The contrast is then null at every width:
$+1.8\%$ at 0.5\,s ($p=0.76$), $+5.1\%$ at 1\,s ($p=0.38$), $+6.4\%$ at 2\,s ($p=0.29$),
$+10.8\%$ at 3\,s ($p=0.11$), $+3.6\%$ at 5\,s ($p=0.61$) and $+17.0\%$ at 10\,s ($p=0.056$; at this width, a fifth of events and five readers are dropped for lack of remaining page). When the window extends to the end of the page, the $+17.5\%$ effect is opportunity, not
repair. It is the same failure as the naive skipping increase in Section~\ref{sec:skip}. In both
cases, the position and time of the events produce the state contrast, not the process that the
measure names. Here the failure comes from the definition of a window, not from the scoring of a
gap.

\paragraph{Interpretation.} The return-rate result is not an artifact of a window that is too
short. No tested window showed a reliable decrease in returns during mind-wandering, which is
the direction that the literature predicts. The result does not show whether late returns behave
differently from immediate ones. The measure that includes them here, the overall backward-step
rate, does not record which word a reader went back to. That question would require a targeting
analysis, which we do not attempt in this paper.

\FloatBarrier
\subsection*{S15. Precision and limits of the coupling contrasts}

\paragraph{Duration equivalence tests.}
We used the stored within-token interactions and reader-clustered standard errors, with a $t$ reference distribution on 43 degrees of freedom. The bounds were $\pm$20\% or $\pm$10\% of each observed on-task slope, treated as a fixed reference. At 20\%, TOST $p$ values were 0.005085 for frequency, 0.001073 for surprisal and 0.002727 for length; Holm-adjusted values were 0.005454, 0.003218 and 0.005454. At 10\%, they were 0.285916, 0.172920 and 0.185899. Thus the three interactions support the 20\% bound but not the 10\% bound. These tests were computed during manuscript preparation and were not preregistered. Uncertainty in the reference slopes is not propagated; the results are conditional on those observed references. The figure's 95\% intervals use the normal-Wald convention of the stored effect summaries.

\begin{table}[!htbp]
\centering
\small
\setlength{\tabcolsep}{4.5pt}
\caption{The coupling contrasts, with what each can and cannot rule out. $\Delta$ is the change
during mind-wandering as a percentage of the on-task effect, positive meaning stronger coupling.
``Attenuation excluded'' is the one-sided 95\% bound, that is, the smallest reduction
inconsistent with the data. The last column is the smallest change the contrast could have
detected at 80\% power. Duration rows are within-token interactions with reader and token fixed
effects, expressed against the on-task slope across words. Return rows take their intervals from
the reader bootstrap. Corrective-return coupling (16 readers) is uninformative and omitted.
Neural rows use the deconvolved kernels from all 44 readers. The two omnibus rows are
model-based and place no restriction on which word properties can be used.}
\label{tab:equiv}
\begin{tabular}{llrrrr}
\toprule
Level & Predictor & Readers & $\Delta$ (\% of on-task) & Attenuation excluded & Detectable at 80\% \\
\midrule
Selection & frequency  & 36 & $+0.5$ $[-3.3, +4.2]$     & $>3\%$  & 6\% \\
Selection & length     & 36 & $+1.8$ $[-1.7, +5.2]$     & $>1\%$  & 5\% \\
Selection & surprisal  & 36 & $+3.1$ $[-2.7, +8.5]$     & $>2\%$  & 8\% \\
Duration  & frequency  & 44 & $+7.3$ $[-1.9, +16.6]$    & $>1\%$  & 13\% \\
Duration  & length     & 44 & $+5.5$ $[-4.2, +15.2]$    & $>3\%$  & 14\% \\
Duration  & surprisal  & 44 & $-5.9$ $[-14.4, +2.6]$    & $>13\%$ & 12\% \\
Refixation & frequency & 35 & $-0.7$ $[-7.9, +6.6]$     & $>7\%$  & 11\% \\
Refixation & length    & 35 & $-2.2$ $[-9.7, +5.1]$     & $>9\%$  & 11\% \\
Refixation & surprisal & 35 & $-3.1$ $[-12.6, +6.1]$    & $>11\%$ & 14\% \\
Regression target & frequency & 38 & $-25.6$ $[-69.2, +14.8]$ & $>62\%$ & 62\% \\
Regression target & length    & 38 & $-27.0$ $[-53.8, -0.6]$  & $>50\%$ & 39\% \\
Regression target & surprisal & 38 & $+0.3$ $[-50.6, +49.6]$  & $>43\%$ & 74\% \\
Omnibus   & target choice & 43 & $-8.9$ $[-17.8, -0.5]$ & $>16\%$ & 12\% \\
Omnibus   & duration      & 43 & $+7.1$ $[-27.2, +42.4]$ & $>21\%$ & 49\% \\
Neural    & frequency  & 44 & $+56.5$ $[-14.3, +143.5]$ & $>4\%$  & 118\% \\
Neural    & surprisal  & 44 & $-35.4$ $[-234.6, +151.5]$ & none   & 283\% \\
\bottomrule
\end{tabular}
\end{table}

\FloatBarrier

\subsection*{S16. Extended methods}

\subsubsection*{Datasets}

\paragraph{Primary sample.} We analyzed the ROAMM dataset \citep{roammdata}: 44 adults read
five Wikipedia articles, each divided into ten pages of about 220 words, rendered as monospaced
text. Reading was self-paced and page-wise, and readers could not return to an earlier page.

The ReMind procedure collected the mind-wandering reports \citep{remind2026}. Readers pressed a
key whenever they caught themselves mind-wandering. They then moved to an annotation screen and
marked the words at which the lapse had begun and ended. We call the marked text the span of the
lapse. Span onset is the first fixation on the reader's marked onset word. Since marked offset words and report times can disagree, the dataset providers define span offset as 2\,s before the
keypress. They exclude the final 2\,s before any keypress, whether a mind-wandering report or
a page advance, to remove motor-preparatory activity. We inherit both conventions.

The reports are self-caught and annotated at the moment of catching, never recalled at the end of
an article. After annotating, readers returned to the same page. The release labels only
first-pass reading, from page onset to the report or page advance. As a result, each page contributes
at most one labeled span, and reading after a report enters none of our analyses. Readers answered
one four-option comprehension question per page after finishing each article.

Eye movements were recorded binocularly with an EyeLink 1000 Plus at 1000\,Hz. EEG was recorded
with a 64-channel BioSemi ActiveTwo system at 2048\,Hz and supplied in the synchronized derivative
at 256\,Hz. Before release the providers re-referenced the EEG to the
common average, band-passed it from 0.5 to 50\,Hz, interpolated bad channels and removed ocular
and muscle components by independent component analysis \citep{roammbench}. The released
derivative provides fixation events mapped to corpus word identifiers. We used all 404{,}557 first-pass fixations; 7.2\% fell inside a reported span.

\paragraph{Positive-control sample.} The control sample is ZuCo 1.0 \citep{hollenstein2018}: 12 adults read
isolated sentences under three tasks, with 105 usable channels of EEG at 500\,Hz and
simultaneous eye tracking. Task 2 is normal reading of encyclopedic sentences, task 3 is a
search for a named semantic relation in encyclopedic sentences, and task 1 is sentiment reading
of film-review sentences. Task order was fixed for every reader: session one contained task 2
followed by the first half of task 1, and session two contained task 3 followed by the second
half of task 1. We exploit that structure to separate task from session, splitting task 1 at
its sentence-index midpoint.

\subsubsection*{Linguistic predictors}

Word frequency was the Zipf value \citep{vanheuven2014,brysbaert2018} from the \texttt{wordfreq}
package \citep{speer2022wordfreq}, which pools subtitle, web, encyclopedia and other English
corpora. Contextual surprisal was the negative log probability of
each word under GPT-2 \citep{radford2019}, in bits, conditioned on the preceding text of its own
sentence. No beginning-of-sequence marker was prepended, so the first word of each sentence has no surprisal
value and is not in the surprisal contrasts. This applies to 656 of 10,839 stimulus words (6.1\%). We used GPT-2 instead of a larger model because surprisal from larger
models fits human reading behavior worse \citep{oh2023}. A word's surprisal is the sum over the
sub-word tokens overlapping its character span (SI~S3). Word length was in letters.

\subsubsection*{Defining the three eye-movement decisions}

We assigned reading positions in corpus order within each article, and for each reader and
article took the ordered sequence of first-pass fixation positions.

\emph{Word skipping.} For each forward step from position $p_i$ to $p_{i+1}>p_i$ we treated the
words strictly between as stepped over. A word counts as skipped if a step spanning at most $G$
intervening words passed over it and no fixation ever landed on it, and as fixated otherwise.
We dropped words that fall only inside steps larger than $G$, because no fixation shows that the
reader scanned them. The primary analysis used $G=4$, and we report results for
$G\in\{1,2,3,4,6,10\}$. The naive alternative counts any unfixated word in the reading span as
skipped and gives it the state of the nearest fixated word. SI~S4 lists every rule. We took the
mind-wandering state of a stepped-over word from the bounding fixations and dropped steps whose
bounding fixations disagreed (3.4\% of steps).

A second criterion makes no reference to gap size: a word
is analyzable if the reader fixated at least one word on the same text line both before and
after it. SI~S4 specifies it alongside the other skipping diagnostics. For the skipping rate we
also report the standard first-pass definition, which counts a word as skipped whenever the
first forward pass stepped over it, even if the reader returned to it later. We applied it with
steps confined to a text line, which leaves out words passed over at a line break, and with
steps allowed to cross lines.

\emph{Gaze duration.} Gaze duration was the sum of first-pass fixations on a word; first-fixation
duration was the first of them. We analyzed the natural logarithm and report raw milliseconds
alongside it where the scale matters.

\emph{Returns to skipped words.} We defined a corrective return for every forward step that stepped over exactly
one word: the event counts as 1 if any of the next five fixations landed on that word. The window is short because returns are fast. Only 23.0\% of stepped-over words receive a return
at all. The median return comes one fixation and 0.30\,s after the step, and five fixations
capture 75.2\% of all returns. SI~S14 repeats the contrast over windows
defined in seconds. In addition, we computed the overall rate of backward steps and the rate of
immediate refixation.

\subsubsection*{Coupling measures and inference}

Two of the three decisions are binary: whether a word was skipped or fixated, and whether a
stepped-over word drew a corrective return. For both we used Somers' $D$, computed as $2A-1$
where $A$ is the probability that a randomly chosen positive case has a higher predictor value
than a randomly chosen negative case \citep{somers1962}. Somers' $D$ is invariant to monotone
transformations of the predictor and to the base rate of the outcome, which matters because the
base rates differ between states. For duration we used ordinary least-squares slopes on
standardized predictors, computed within reader and within state.

Reader is the unit of inference throughout. We report the mean, a 95\% percentile interval from
10{,}000 reader-level bootstrap resamples, a paired $t$ test, and the number of readers showing
the effect. We corrected families of three tests by the Holm procedure \citep{holm1979}.
Retention is the mind-wandering estimate divided by the on-task estimate, computed on signed
values. We supported each null with two one-sided tests against bounds of 10\% and 20\% of the
on-task effect \citep{lakens2017} and with default Bayes factors for the paired difference with
a Cauchy prior of scale $1/\sqrt{2}$ \citep{rouder2009}.

\subsubsection*{Within-token fixed-effects model}

Because the same corpus token is read by many readers, mind-wandering status varies within
token. We fitted
\[
\log(\text{gaze duration})_{ij}
= \alpha_i + \gamma_j + \delta\,m_{ij} + \sum_k \theta_k\,m_{ij}z_{k,i} + \varepsilon_{ij},
\]
where $i$ indexes token instance, $j$ indexes reader, $m$ is the mind-wandering indicator and
$z_k$ are standardized word properties. An alternating within-transformation absorbed the token
and reader effects, and we clustered standard errors on reader \citep{cameron2015}. The
interaction then comes purely from readers who differ in state on the same physical word in the
same position of the same page.

\subsubsection*{Fixation-related potentials and deconvolution}

Starting from the providers' cleaned recording, we re-referenced each run to the average and
applied a 0.1 to 30\,Hz zero-phase band pass. In natural reading, the responses to successive fixations overlap. The overlap is correlated with word properties, since readers fixate difficult words longer. For this reason, naive fixation-locked averages are not interpretable \citep{dimigen2011,dimigen2021}. We instead estimated overlap-corrected kernels
by linear deconvolution \citep{smith2015,ehinger2019}. For each reader, we built a sparse finite-impulse-response design over $-100$ to $+500$\,ms around
every fixation onset. It contained an intercept, frequency, length, surprisal and log fixation
duration, a mind-wandering indicator, and the interactions of frequency and surprisal with that
indicator. The intercept kernel is the on-task fixation-related response. The indicator kernel is
its change during mind-wandering. The frequency and surprisal kernels are the lexical and semantic
couplings, and their interactions are the change in those couplings.

We solved the normal equations with a ridge penalty proportional to each column's own energy,
$\lambda_j = 10^{-2}\,(X^\top X)_{jj}$, so that every predictor is shrunk by the same small
fraction. The choice matters for the state kernels. The mind-wandering columns are nonzero on 7\% of fixations. A single penalty sized to the dense
columns shrinks them several times more in relative terms, which biases the change during
mind-wandering toward zero. We report this penalty as a sensitivity comparison. SI~S7 reports
every state estimate under both schemes and four penalty strengths.

We analyzed the two regions and time windows established for these responses in natural reading:
an occipitotemporal set from 150 to 290\,ms for the frequency response and a centroparietal set
from 300 to 450\,ms for the N400 \citep{dimigen2011,kutas2011,frank2015}. The same two windows
serve every contrast in the paper, including the rescaling test. Beyond those
regions we ran cluster-based permutation tests over the full epoch and all channels with
sign-flipping at the reader level \citep{maris2007}. We also asked whether the change during mind-wandering alters the pattern or only the size of the
fixation-related response. We estimated a single rescaling factor that relates the difference
field (mind-wandering minus on-task) to the on-task field over 150 to 290\,ms, with
leave-one-reader-out estimation. We then tested the residual field with a sign-flip test on global
field power (SI~S7).

The early fixation-related response depends on the saccade that brings the eyes to a word, and
independent component analysis leaves some ocular activity in free viewing. We therefore refitted
the model with the incoming saccade as a covariate, following \citet{dimigen2021}: a cubic spline
of log saccade amplitude, the signed horizontal and vertical displacement, and indicators for
the saccade type (forward within a line, regression, refixation, sweep to the next line, return
to an earlier line). Finally, we repeated the rescaling test on forward saccades matched in amplitude across states, in
0.25$^\circ$ bins within reader. We averaged over ten matching draws and compared the result with
random subsets of the same size (SI~S7).

\subsubsection*{Ongoing band power}

We cut the continuous average-referenced recording into non-overlapping 2\,s segments that were
wholly inside one state. We computed log power in the theta (4--8\,Hz), alpha (8--13\,Hz), beta
(13--30\,Hz) and gamma (30--45\,Hz) bands from a Hann-windowed periodogram. Per reader and per
channel we compared the two states with a Welch $t$ test and expressed the result as a
standardized difference. Across readers we tested every channel with a sign-flip max-$t$
procedure that controls the family-wise error over all 64 channels. A clean variant dropped every
segment containing a blink and excluded from on-task the 10\,s after each span offset, which
covers the report and annotation screens.

To separate an ongoing oscillation from the evoked response, we repeated the segment analysis
after subtracting each reader's deconvolved response to every fixation (SI~S9). A reader entered a
contrast with at least five complete segments in each state. One reader had too little
mind-wandering for either variant, and two more fell below five segments once blink segments
were dropped.

\subsubsection*{Comprehension outcomes and answer spans}

The released dataset records one four-option comprehension question per page. We parsed 2{,}200
per-page outcomes and validated them against the released accuracy fields and against the item
keys. One item's key contradicts its page. We retain it in the models reported here, and report its exclusion as a sensitivity analysis; excluding it changes no principal conclusion (SI~S12). The
correspondence between the physiology subject index and the released subject identifiers is not stated in the release, and we derived it from the pattern of reading durations across the 50 pages. The recovered mapping
is bijective and agrees with sorted identifier order for all 44 readers.

We localized each item to its answer-bearing sentences twice. Every result reported here uses the primary annotation. An AI model (Claude Opus 5) made it: the
model read each page and item in context and marked the sentences that a reader must have encoded
to choose the keyed answer. For items phrased with a negation, the evidence is the union of the
sources of the true options.
The resulting sentence indices and the rubric are in the code repository. The second annotation scored every sentence on the page against the question and its options by
sentence embedding and inverse-document-frequency overlap. We corrected seven items by hand where
a fact splits across sentences. SI~S12 reports their agreement and the checks on each.

A trial entered the models if the reader made at least three fixations on the answer span and
ten elsewhere on the page. This keeps 1804 of the 2200 trials, and 326 answer spans were never
fixated. An answer of ``not sure'' counts as incorrect. The lapse measure for a region is the fraction of the reader's fixations on it that fell inside a
reported span. The gaze measure for a region is the fraction of its words that the reader fixated. Effects are in percentage points of
accuracy, either as the effect of any lapse on the region or as the change from none of the
region to all of it (lapsed, or fixated). Regions are compared with one another on the second
scale.

Trial-level models are linear probability models absorbing reader and item, with standard
errors clustered by reader and normal-Wald confidence intervals from reader-clustered standard errors. The
random-set control draws 1000 sentence-sampled text sets of the same word count from the same page, truncating the final sentence where needed
and repeats the analysis on each. The overlap gradient regresses the effect obtained for each
random region on that region's word overlap with the true answer span. We tested alpha with the same design. For each fixation, we took the mean log 8--13\,Hz Hilbert
power from 500\,ms before to 500\,ms after onset over the left posterior-lateral cluster and z-scored it within run. The one-second window was clipped at recording-run boundaries.
We averaged this value over the fixations on the answer span. It entered the model alongside the
mind-wandering terms and was compared with a random-set null of its own.

\subsubsection*{Alignment decomposition}

To relate the rescaling to the encoding-model result of \citet{sun2026alignment}, we fitted a
ridge encoder from GPT-2 hidden states to fixation-related amplitude and split the between-state
ratio of encoding accuracy exactly into a response-gain, a stimulus and a noise term (SI~S11).

\subsubsection*{Software and order of analysis}

We ran the analyses in Python with NumPy, SciPy, pandas, scikit-learn, MNE and PyTorch. Dataset and code access are described in the main manuscript's availability statements.

This study was not registered, and the order in which we ran the analyses affects how they should
be interpreted. The duration contrast came first. The skipping and return channels were then
written down, with their hypotheses and exclusion rules, before any of their contrasts were
computed. We did not anticipate that the skipping marker depends on its definition. This result came from an
audit of the skip definition, and we report it as a specification curve (Section~\ref{sec:skip}). The answer-span
annotations were written before any outcome model was fitted; the localization result was thus not tuned against the outcome it predicts. The proportional rescaling, the alignment decomposition,
the band-power analysis and the time course of the rescaling were all prompted by earlier results. For the induced-power test we set the criterion at 70\%
retention before running it. The
column-wise ridge penalty was selected after diagnosing disproportionate shrinkage of sparse state columns; SI~S7 reports estimates under both scaling schemes. The saccade covariates, the
first-pass skipping rule, the skipping analyses of the control sample and its separation of task
from materials were exploratory additions to the analysis.


\end{bibunit}
\end{document}